\documentclass[journal]{IEEEtran}

\usepackage[numbers,sort&compress]{natbib}
\usepackage{url}
\usepackage{bm}
\usepackage{amsfonts}
\usepackage{amsmath}
\usepackage{amssymb}
\usepackage{subfig}
\usepackage{array,color}
\usepackage{tabularx}
\usepackage{multirow}
\usepackage{booktabs}
\usepackage{amsmath}  
\usepackage{makecell}
\usepackage{graphicx}
\usepackage{epstopdf}
\usepackage{graphics}
\usepackage{epsfig}
\usepackage{psfrag}

\begin{document}

\newcommand {\Data} [1]{\mbox{${#1}$}}  

\newcommand {\DataN} [2]{\Data{\Power{{#1}}{{|{#2}}}}}  
\newcommand {\DataIJ} [3]{\Data{\Power{#1}{{|{#2}\!\times{}\!{#3}}}}}  

\newcommand {\DatasI} [2]{\Data{\Index{#1}{\Data 1},\Index{#1}{\Data 2},\cdots,\Index{#1}{#2},\cdots}}   
\newcommand {\DatasII} [3]{\Data{\Index{#1}{{\Index{#2}{\Data 1}}},\Index{#1}{{\Index{#2}{\Data 2}}},\cdots,\Index{#1}{{\Index{#2}{#3}}},\cdots}}  

\newcommand {\DatasNTt}[3]{\Data{\Index{#1}{{#2}{\Data 1}},\Index{#1}{{#2}{\Data 2}},\cdots,\Index{#1}{{#2}{#3}}} } 
\newcommand {\DatasNTn}[3]{\Data{\Index{#1}{{\Data 1}{#3}},\Index{#1}{{\Data 2}{#3}},\cdots,\Index{#1}{{#2}{#3}}} } 

\newcommand {\Vector} [1]{\Data {\mathbf {#1}}}
\newcommand {\Rdata} [1]{\Data {\hat {#1}}}
\newcommand {\Tdata} [1]{\Data {\tilde {#1}}} 
\newcommand {\Udata} [1]{\Data {\overline {#1}}} 
\newcommand {\Fdata} [1]{\Data {\mathbb {#1}}} 
\newcommand {\Prod} [2]{\Data {\prod_{\SI {#1}}^{\SI {#2}}}}  
\newcommand {\Sum} [2]{\Data {\sum_{\SI {#1}}^{\SI {#2}}}}   
\newcommand {\Belong} [2]{\Data{ {#1} \in{}{#2}}}  

\newcommand {\Abs} [1]{\Data{ \lvert {#1} \rvert}}  
\newcommand {\Mul} [2]{\Data{ {#1} \times {#2}}}  
\newcommand {\Muls} [2]{\Data{ {#1} \! \times \!{#2}}}  
\newcommand {\Div} [2]{\Data{ \frac{#1}{#2}}}  
\newcommand {\Trend} [2]{\Data{ {#1}\rightarrow{#2}}}  
\newcommand {\Sqrt} [1]{\Data {\sqrt {#1}}} 
\newcommand {\Sqrtn} [2]{\Data {\sqrt[2]{#1}}} 

\newcommand {\Power} [2]{\Data{ {#1}^{\TI {#2}}}}  
\newcommand {\Index} [2]{\Data{ {#1}_{\TI {#2}}}}  

\newcommand {\Equ} [2]{\Data{ {#1} = {#2}}}  
\newcommand {\Equs} [2]{\Data{ {#1}\! =\! {#2}}}  
\newcommand {\Equu} [2]{\Data{ {#1} \equiv {#2}}}  

\newcommand {\LE}[0] {\leqslant}
\newcommand {\GE}[0] {\geqslant}
\newcommand {\INF}[0] {\infty}
\newcommand {\MIN}[0] {\min}
\newcommand {\MAX}[0] {\max}

\newcommand {\Funcfx} [2]{\Data{ {#1}({#2})}}  
\newcommand {\Funcfzx} [3]{\Data{ {\Index {#1}{#2}}({#3})}}  
\newcommand {\Funcfnzx} [4]{\Data{ {\Index {\Power{#1}{#2}}{#3}}({#4})}}  
\newcommand {\SI}[1] {\small{#1}}
\newcommand {\TI}[1] {\tiny {#1}}
\newcommand {\Text}[1] {\text {#1}}

\newcommand {\VtS}[0]{\Index {t}{\Text {s}}}
\newcommand {\Vti}[0]{\Index {t}{i}}
\newcommand {\Vt}[0]{\Data {t}}
\newcommand {\VMSR}[0]{\Index {\kappa} {\SI{\Index {}{ \Text{MSR}}}}}
\newcommand {\VAT}[0]{\Index {\Vector A}{\Text{Time}}}
\newcommand {\VPbus}[1]{\Index {P}{\Text{Bus-}{#1}}}

\newcommand {\VtSS}[2]{\Equs {\Index {t}{\Text {s}}}{#1} {#2}}
\newcommand {\Vtii}[2]{\Equs {\Index {t}{i}}{#1} {#2}}
\newcommand {\Vtt}[2]{\Equs {t}{#1} {#2}}
\newcommand {\VMSRR}[2]{\Equs {\Index {\kappa} {\SI{\Index {}{ \Text{MSR}}}}}{#1} {#2}}
\newcommand {\VATT}[2]{\Equs {\Index {\Vector A}{\Text{Time}}}{#1} {#2}}
\newcommand {\VPbuss}[3]{\Equs {\Index {P}{\Text{Bus-}{#1}}}{#2} \Text{ #3}}

\newcommand {\VV}[1]{\Index {\Vector V}{#1}}
\newcommand {\VX}[1]{\Index {\Vector X}{#1}}
\newcommand {\Vx}[1]{\Index {\Vector x}{\SI{\Index {}{#1}}}}
\newcommand {\Vsx}[1]{\Index {x}{\SI{\Index {}{#1}}}}
\newcommand {\VZ}[1]{\Index {\Vector Z}{#1}}
\newcommand {\Vz}[1]{\Index {\Vector z}{\SI{\Index {}{#1}}}}
\newcommand {\Vsz}[1]{\Index {z}{\SI{\Index {}{#1}}}}
\newcommand {\VRX}[1]{\Index {\Rdata {\Vector X}}{#1}}
\newcommand {\VRx}[1]{\Index {\Rdata {\Vector x}}{\SI{\Index {}{#1}}}}
\newcommand {\VRsx}[1]{\Index {\Rdata {x}}{\SI{\Index {}{#1}}}}
\newcommand {\VRZ}[1]{\Index {\Rdata {\Vector Z}}{#1}}
\newcommand {\VRz}[1]{\Index {\Rdata {\Vector z}}{\SI{\Index {}{#1}}}}
\newcommand {\VRsz}[1]{\Index {\Rdata {z}}{\SI{\Index {}{#1}}}}
\newcommand {\VTX}[1]{\Index {\Tdata {\Vector X}}{#1}}
\newcommand {\VTx}[1]{\Index {\Tdata {\Vector x}}{\SI{\Index {}{#1}}}}
\newcommand {\VTsx}[1]{\Index {\Tdata {x}}{\SI{\Index {}{#1}}}}
\newcommand {\VTZ}[1]{\Index {\Tdata {\Vector Z}}{#1}}
\newcommand {\VTz}[1]{\Index {\Tdata {\Vector z}}{\SI{\Index {}{#1}}}}
\newcommand {\VTsz}[1]{\Index {\Tdata {z}}{\SI{\Index {}{#1}}}}
\newcommand {\VOG}[1]{\Vector{\Omega}{#1}}

\newcommand {\Sigg}[1]{\Data {\sigma}^2({#1})}
\newcommand {\Sig}[1]{\Data {\sigma}({#1})}
\newcommand {\Mu}[1]{\Data {\mu} ({#1})}
\newcommand {\Eig}[1]{\Data {\lambda}({\Vector {#1}}) }
\newcommand {\Her}[1]{\Power {#1}{\!H}}
\newcommand {\Tra}[1]{\Power {#1}{\!T}}

\newcommand {\VF}[3] {\DataIJ {\Fdata {#1}}{#2}{#3}}

\newcommand {\Tcol}[2] {\multicolumn{1}{#1}{#2} }
\newcommand {\Tcols}[3] {\multicolumn{#1}{#2}{#3} }
\newcommand {\Cur}[2] {\mbox {\Data {#1}-\Data {#2}}}

\def \FuncC #1#2{
\begin{equation}
{#2}
\label {#1}
\end{equation}
}

\def \FuncCC #1#2#3#4#5#6{
\begin{equation}
#2=
\begin{cases}
    #3 & #4 \\
    #5 & #6
\end{cases}
\label{#1}
\end{equation}
}

\def \Figff #1#2#3#4#5#6#7{   
\begin{figure}[#7]
\centering
\subfloat[#2]{
\label{#1a}
\includegraphics[width=0.23\textwidth]{#4}
}
\subfloat[#3]{
\label{#1b}
\includegraphics[width=0.23\textwidth]{#5}
}
\caption{\small #6}
\label{#1}
\end{figure}
}

\def \Figffb #1#2#3#4#5#6#7#8#9{   
\begin{figure}[#9]
\centering
\subfloat[#2]{
\label{#1a}
\includegraphics[width=0.23\textwidth]{#5}
}
\subfloat[#3]{
\label{#1b}
\includegraphics[width=0.23\textwidth]{#6}
}

\subfloat[{#4}]{
\label{#1c}
\includegraphics[width=0.48\textwidth]{#7}
}
\caption{\small #8}
\label{#1}
\end{figure}
}

\def \Figffp #1#2#3#4#5#6#7{   
\begin{figure*}[#7]
\centering
\subfloat[#2]{
\label{#1a}
\begin{minipage}[t]{0.24\textwidth}
\centering
\includegraphics[width=1\textwidth]{#4}
\end{minipage}
}
\subfloat[#3]{
\label{#1b}
\begin{minipage}[t]{0.24\textwidth}
\centering
\includegraphics[width=1\textwidth]{#5}
\end{minipage}
}
\caption{\small #6}
\label{#1}
\end{figure*}
}

\def \Figf #1#2#3#4{   
\begin{figure}[#4]
\centering
\includegraphics[width=0.48\textwidth]{#2}

\caption{\small #3}
\label{#1}
\end{figure}
}

\title{3D Power-map for Smart Grids---An Integration of High-dimensional Analysis and Visualization}
%
%
%

\author{Xing~He, Qian~Ai, ~\IEEEmembership{Member,~IEEE,}, Robert~C. Qiu,~\IEEEmembership{Fellow,~IEEE,},
Jianmo~Ni, Longjian~Piao, Yiting~Xu, Xinyi~Xu
\thanks{This work was supported by National Key Technology Research and Development Program of Science and Technology (2013BAA01B04).}
\thanks{Xing~He, Qian~Ai, Jianmo~Ni, Longjian~Piao,  Yiting~Xu, Xinyi~Xu are with the Department of Electrical Engineering, Shanghai Jiaotong University, Shanghai 200240, China (e-mail: {hexing\_hx@126.com;\- aiqian@sjtu.edu.cn)}}
\thanks{Robert~C.~Qiu is with the Department of Electrical and Computer Engineering,
Tennessee Technological University, Cookeville, TN 38505 USA (e-mail: {rqiu@tntech.edu})}}

\maketitle

\begin{abstract}
Data with features of volume, velocity, variety, and veracity are challenging traditional tools to extract useful analysis for decision-making. By integrating high-dimensional analysis with visualization, this paper develops a 3D power-map animation as an effective solution to the challenge. An architecture design, with detailed data processing procedure, is proposed to realize the integration. Two of the most important components in the architecture are presented: the Single-Ring Law for random matrices as solid mathematic foundation, and the proposed statistic MSR as high-dimensional data for visualization. The whole procedure is easy in logic, fast in speed, objective and even robust against bad data. Moreover, it is an unsupervised machine learning mechanism directly oriented to the raw data rather than logics or models based on simplifications and assumptions. A case study validates the effectiveness and performance of the developed 3D power-map in analysis extraction.
\end{abstract}


%
\IEEEpeerreviewmaketitle

\section{Introduction}

%
\IEEEPARstart{B}{ig} Data analysis and its visualization are efficient solutions to data explosion in smart grids \cite{qiu2014smart,kezunovic2013role,pao1998visualization,dos2004gaining}. The rapid development of power systems leads to more challenging datasets \cite{IBM2009Manag}. For the datasets, we can hardly handle data with features of volume, velocity, variety, and veracity (i.e. 4Vs data) \cite{IBM2014fourv} within a tolerable elapsed time or hardware resources. This challenge has encouraged the development of an emerging and interesting field---big data analysis and its visualization. This field aims to extract analysis directly from the multivariate and multidimensional raw data in high-dimensional perspectives. In other words, big data high-dimensional analysis and its visualization establish and illustrate the inherent correlations among the raw data to help understand the system and to gain insight to the mechanism, rather than build and analyze the models based on simplifications and assumptions \cite{qiu2013bookcogsen}.

It is generally believed that in the coming century, big data high-dimensional analysis will be a very significant activity \cite{donoho2000aide,nature2008bigd,science2011bigd}. It has been successfully applied as a powerful tool in specific fields for numerous physical phenomena, such as quantum \cite{brody1981random}, financial systems \cite{laloux2000random}, biological systems \cite{howe2008big}, as well as wireless communication networks \cite{qiu2014Intial70N,qiu2014MIMO,qiu2013bookcogsen}. It is obvious and significant that big data high-dimensional analysis applications will appear in smart grids as in other fields.
Visualization is an essential issue in power system operation, forecasting, fault detection, and even Economic Dispatch. It is a key tool to extract analysis from the overwhelming amount of data to support decision-making \cite{xu20063d}. Much work has been previously done in the area of developing visualization techniques to aid in interpreting power system data. In \mbox{\cite{tang2011smart,yan2011design,zhu2011data,hu2011research}}, two kinds of visualization methods are used to visualize the state variables in the system, which are voltage magnitudes and phase angles for all the buses. Color contouring is used for voltage magnitude while the varying height of the underneath terrain represents the phase angles. The 4Vs data in smart grids, however, are hard to be visualized with traditional tools. One of the major challenges is to find an effective way to convert the high-dimensional data to low-dimensional geometry as a man-machine interface.

This paper integrates high-dimensional analysis with visualization to develop a 3D power-map animation. Firstly, the single-ring law for random matrix is introduced as mathematical foundations. Then an architecture is proposed as a universal solution to realize the integration. In the architecture, Mean Spectral energy Radius (MSR) is defined as a new high-dimensional statistic to visualize the data correlations. In addition, MSR clarifies the information which should be interchanged among each distributed partition for analysis extraction. A case study is presented to show the effectiveness and performance of the developed 3D power-map with the proposed statistic MSR.

\section{Mathematical Theory and Data Procedure}
\subsection{The Single-Ring Law}
Consider the matrices product \Data {\VTZ{}=\Prod{i=1}{L} \VX {u,i}}, where \Belong{\VX u}{\VF CNN} is the singular value equivalent \cite{ipsen2014weak} of the rectangular \Muls {N}{T}  non-Hermitian random matrix \VTX{},
whose entries are independent identically distributed (i.i.d.) variables with mean \Equs {\Mu{\VTx{k,:}}}{0}  and  variance \Equs {\Sigg{\VTx{k,:}}}{1} for \Equs{k}{\Data {(1,2,\cdots,N)}}. The matrices product \VTZ{}  can be converted to \VZ {} by a transform which make the variance to \Equs {\Sigg{\VTz{:,k}}}{1/N} for \Equs{k}{\Data {(1,2,\cdots,N)}}. Thus, the empirical eigenvalues distribution of \VZ {} converges almost surely to the same limit given by
\FuncCC {eq:Lambda}
{\Funcfzx {f}{\VZ {}}{\lambda}}
{ \Div{1}{\pi{}c\alpha}{\Power{\Abs{\lambda}}{(2/\alpha-2)}}}  {{\Text {, }} \Power {(1-c)}{\alpha/2} \LE \Abs{\lambda} \LE \Data 1   }
{0}                                   {\Text {, otherwise}}
as \Trend {N,T}{\INF} with the ratio \Equ {N,T}{\Belong {c} {(0,1]} }.  On the complex plane of the eigenvalues, the inner circle radius \Index {r}{\MIN \_\Eig Z}  is \Power{(1-c)}{\alpha/2} and outer circle radius is unity. In addition, we propose the mean spectral energy radius as an statistic MSR to characterize the distribution. Figure \ref{fig:SingleRL} shows the cases of \Data {L=1 } and \Data {L=8}, respectively.

\Figff {fig:SingleRL} {\Data {L=1}}{\Data {L=8}}{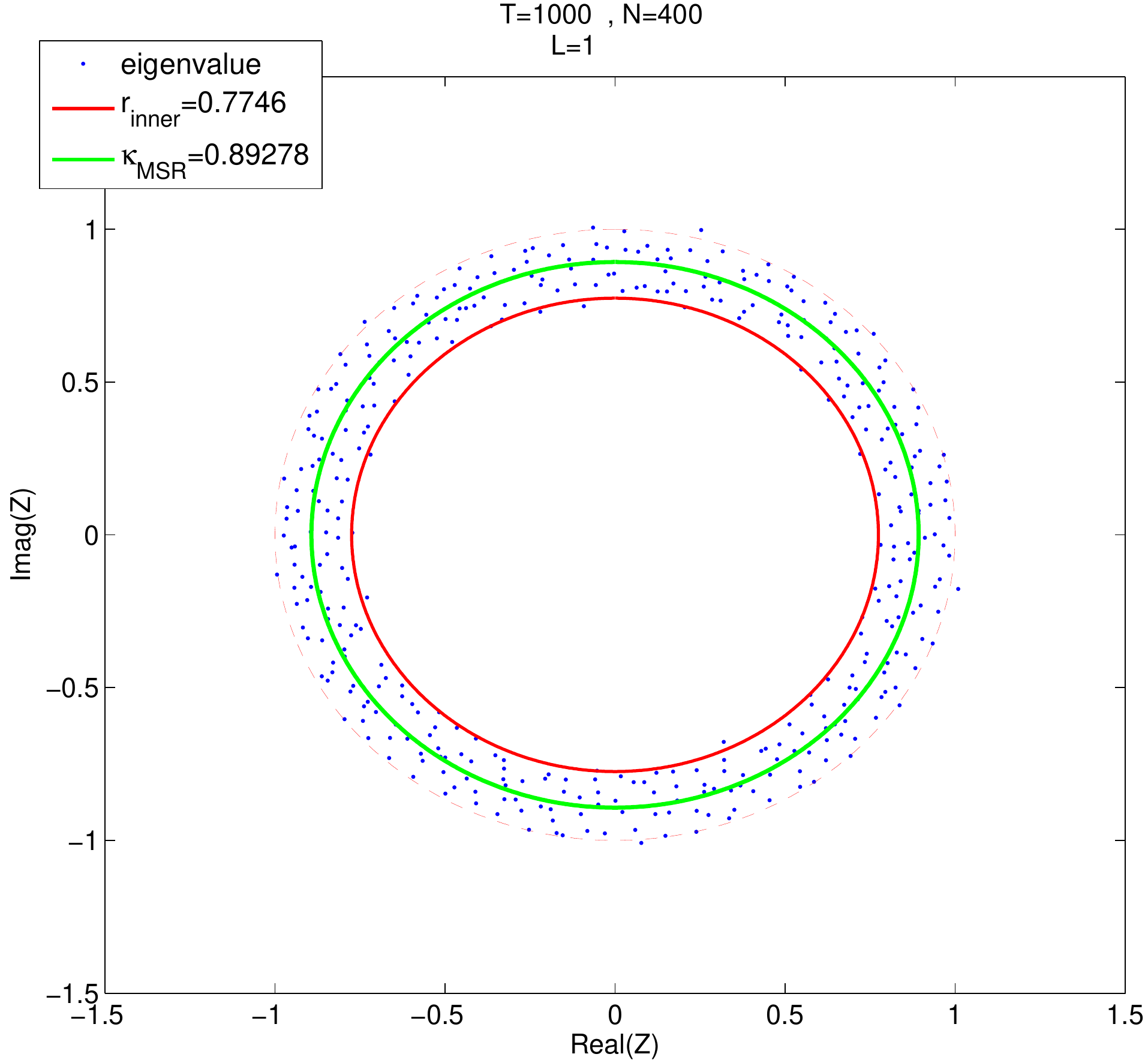}{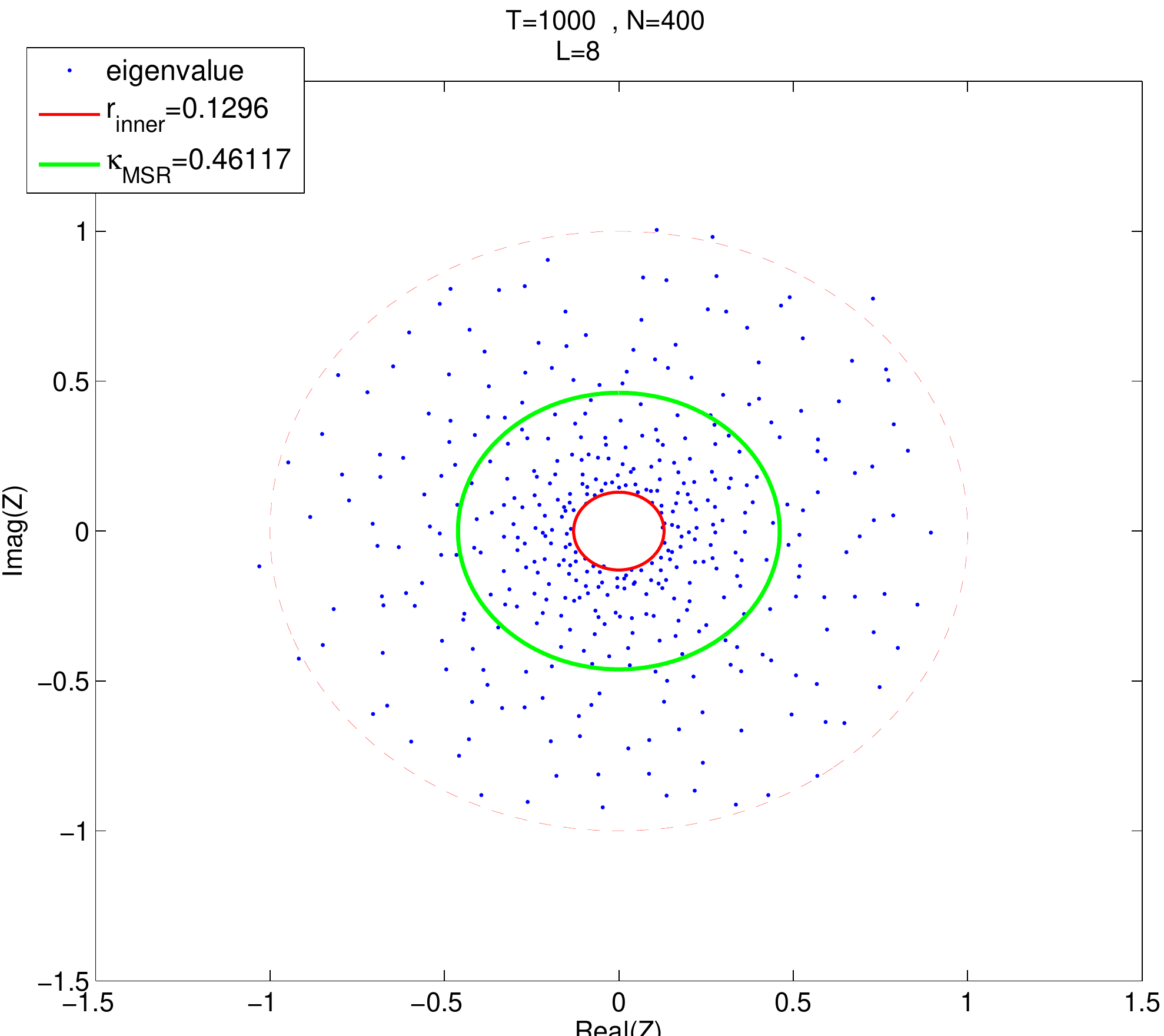}
{Eigenvalues \Eig Z   for a product of \Data L non-Hermitian random matrices for \Data {N=400}, \Data {T=1000}. Figure (a) is for \Data {L=1}, and (b) is for \Data {L=8}. The red and green line show the inner radius and the mean spectral energy radius, respectively.}
{htbp} 


\subsection{From Power Systems to Single-Ring Law}

For a power system, at a certain time \Vti, the raw data \VRsx{}  can be arranged to a vector  \VRx{\Vti}. As time goes by, vectors are acquired one by one and a sheet as data set is naturally formed to map the systems. Any arbitrary section in this set can be selected as raw data source \VOG {\VRx{}}  for further analyses. Thus, the \VOG {\VRx{}}  consists of sample vectors on a series of times denoted as \DatasII {\VRx{}} {t} {i};  at any time \Vti, the vector \VRx {\Vti} comprises sample data denoted as \DatasII {\VRsx{}} {t_i,n}{k}.  The length of \Data {n} is decided by the varieties of data at a single sampling time, and the length of \Data {t} is subject to the volume of historical data set and is generally big enough.

For the raw data source \VOG {\VRx{}}, we can focus on any data area as a split-window to form a raw matrix \VRX{}.  Then, we convert it to a standard rectangular non-Hermitian matrix  \VTX{}. It is performed by rows with following algorithms:
\FuncC {eq:StdMatrix}
{\small{
\VTsx{ij}\!=\!\Muls{  (\VRsx{ij}\!-\!\Udata {\VRx{i}}) } {( \Sig{\VTx{i}}/\Sig{\VRx{i}} )\! + \! \Udata {\VTx{i}}}
\,,\Data {1\LE{}i\LE{}N\!;1\LE{}j\LE{}T}}
}
where \Equs {\VRx{i}}{(\DatasNTt {\VRsx{}}{i}{T})}  and \Equs {\Udata {\VTx{i}}}{0}, \Equs {\Sigg {\VTx{i}}}{1}.

The  matrix \Belong {\VX {u}} {\VF CNN} is introduced as the singular value equivalent of the matrix
 \Belong {\VTX{}} {\VF CNT} by
\FuncC {eq:Xu}
{\Equ {\VX u}   {\Vector U \Sqrt{\VTX{}{\Her{\VTX{}}}}}}
where \Belong {\Vector U} {\VF CNN} and \Equu {\Her{\VX{u}}\VX{u}} {\VTX{}{\Her{\VTX{}}}}.

Then, the matrices product \Equ {\VTZ {}}{ \Prod{i=1}{L}{\VX {u,i}}}  is acquired, based on which, \VZ{} is calculated by the rows with following formula:
\FuncC {eq:Z}{
{\Equ {\Vz{j}} {\VTz{j}/({\Sqrt N}\Sig{\VTz{j}})}}
\quad ,\Data {1\LE{}j\LE{}N}
}
where \Equs {\Vz{j}}{(\DatasNTn {\Vsz{}}{N}{j})^T } , \Equs {\VTz{j}}{(\DatasNTn {\VTsz {}}{N}{j})^T }

The above procedure of variable transformation is depicted as follows. The matrix \VZ{} is calculated for single-ring law and the mean value of all the eigenvalues' radii \VMSR{}  is proposed as a new high-dimensional statistic.

\begin{table}[htbp]
\centering

\begin{tabular*}{8.8cm} {l}
\toprule[1.5pt]
\textbf {Steps of Variables Transformation} \\
\toprule[0.5pt]
{ }{ }1) raw matrix \Belong {\VRX{}}{\VF CNT}\\
{ }{ }2) standard rectangular matrix \Belong {\VTX{}}{\VF CNT}, \Equs {\Mu{\VTx{k,:}}}{0}, \Equs {\Sigg {\VTx{k,:}}}{1}  \\
{ }{ }3) singular value equivalent \Equ { \VX u}{\Vector U\Sqrt {\VTX{}\Her{\VTX{}}}}\Belong {}{\VF CNN}\\
{ }{ }4) matrices product \Equ {\VTZ {}}  {\Prod {i=1}{L}{\VX{u,i} }} \Belong {}{\VF CNN}    \\
{ }{ }5) \Belong {\VZ{}}{\VF CNN},  \Equs {\Mu {\VTz{:,k}}}{0}, \Equs {\Sigg{\VTz{:,k}}}{1/N}\\
{ }{ }6) \Equ {\VMSR} {\Udata {\Index {\Vector r}{\Eig Z}}} \\

\toprule[1pt]

\end{tabular*}

\end{table}

\subsection{Universal Architecture and its Advantages}
Based on the Single-ring Law and above transform variables, we design following steps as the general architecture to conduct analysis and visualization in this paper.

\begin{table}[htbp]
\centering

\begin{tabular*}{8.8cm} {l}
\toprule[1.5pt]
\textbf {Steps of High-dimensional Analysis and its Visualizations} \\
\toprule[0.5pt]
{ }{ }1) Form \Vector Z for any time \Index {t}{i} as described in Section II\\
{ }{ }2) Calculate eigenvalues \Eig Z\\
{ }{ }3) Plot the single-ring\\
{ }{ }4) Conduct high-dimensional analysis\\
{ }{ }{ }{ }{ }{ }4a) Observe and Compare the results with standard ring\\
{ }{ }{ }{ }{ }{ }4b) Calculate the statistic \VMSR{}\\
{ }{ }5) Conduct visualization\\
{ }{ }6) Conduct interpretations\\

\toprule[1pt]
\end{tabular*}
\end{table}

With a pure statistical procedure, steps 1)--4) of the proposed architecture \cite{he2015arch} enable us to conduct high-dimensional analysis of the interrelation and interaction among the raw data seen as “correlations”. Based on the correlation, we try to extract analysis directly from the raw data without simplifications and assumptions for engineering model or causal logic. As a result, the procedure is easier in logic, faster in speed, and objective without introducing or accumulating errors. Moreover, it is an unsupervised machine learning mechanism \cite{he2015unsup} directly oriented to a matrix consisting of 4Vs raw data mentioned in Section I. In addition, due to the high-dimensional feature, the analysis is more robust against the incomplete, inaccurate, and unavailable data. These advantages will be detailed in the case study.

\section{Case Study}
Figure \ref{fig:case118} shows the standard IEEE 118-bus system with six partitions \cite{ni2007new}. Detailed information about the test bed is referred to the \emph {case118.m} in Matpower package and Matpower 4.1 User’s Manual \cite{MATPOWER2011matpower}. Table I is for this case.

\Figf {fig:case118} {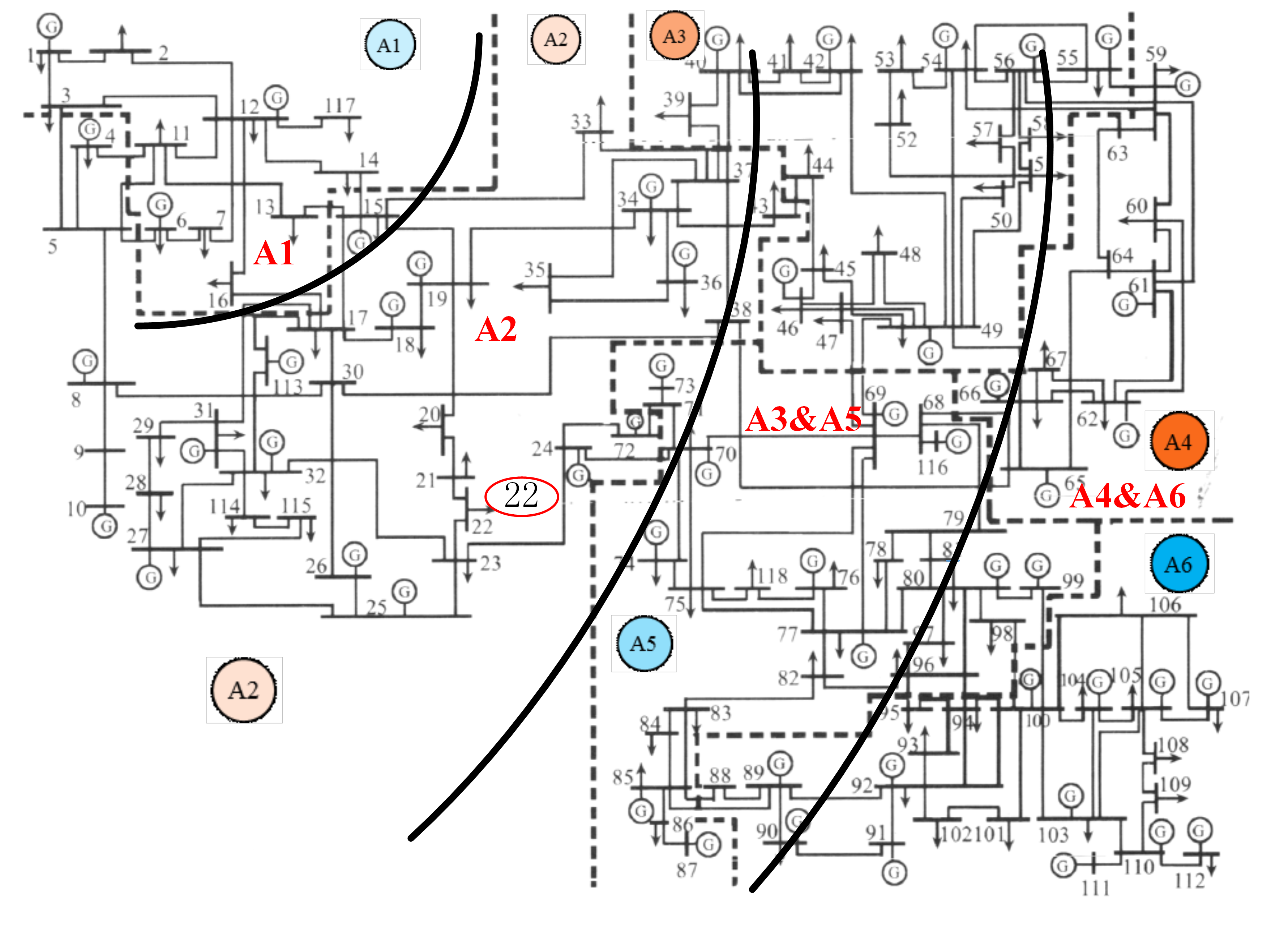}
{Partitioning network for IEEE 118-bus system. There are six partitions, i.e. A1, A2, A3, A4, A5, and A6.}
{htbp} 

There are generally two scenarios in this case: 1) only white noses, such as small random fluctuations of loads or sample errors; 2) Signals plus noises, such as sudden changes or faults at certain bus or the grid.
Table \ref{Tab: Event Series} shows the series of assumed events. The distribution of eigenvalues \Eig {Z} in the single-ring at \VtSS{300}{s}, \VtSS{301}{s}, and the MSR on the time series \VMSR{}  are illustrated by Figure \ref{fig:Case Ringa}, \ref{fig:Case Ringb}, and \ref{fig:Case Ringc}, respectively.

\begin{table}[htbp]
\caption {Series of Events}
\label{Tab: Event Series}
\centering

\begin{minipage}[!t]{0.48\textwidth}
\centering

\begin{tabularx}{\textwidth} { >{\scshape}l !{\color{black}\vrule width1pt}    >{$}l<{$}    >{$}l<{$}   >{$}l<{$}  }  
\toprule[1.5pt]
\hline
\textbf {Par}$\backslash$\VAT{} & [001:300] & [301:700] & [701:1000]\\
\hline
\VPbus{22} (MW) & 0 & 200 & t-500\\

\toprule[1pt]
\end{tabularx}
\raggedright
\small {*\VPbus{22} is the power demand of bus-22}
\end{minipage}
\end{table}

\Figffb {fig:Case Ring}
{Sampling time \VtSS{300}{s}}{Sampling time \VtSS{301}{s}}{MSR on Time Series}
{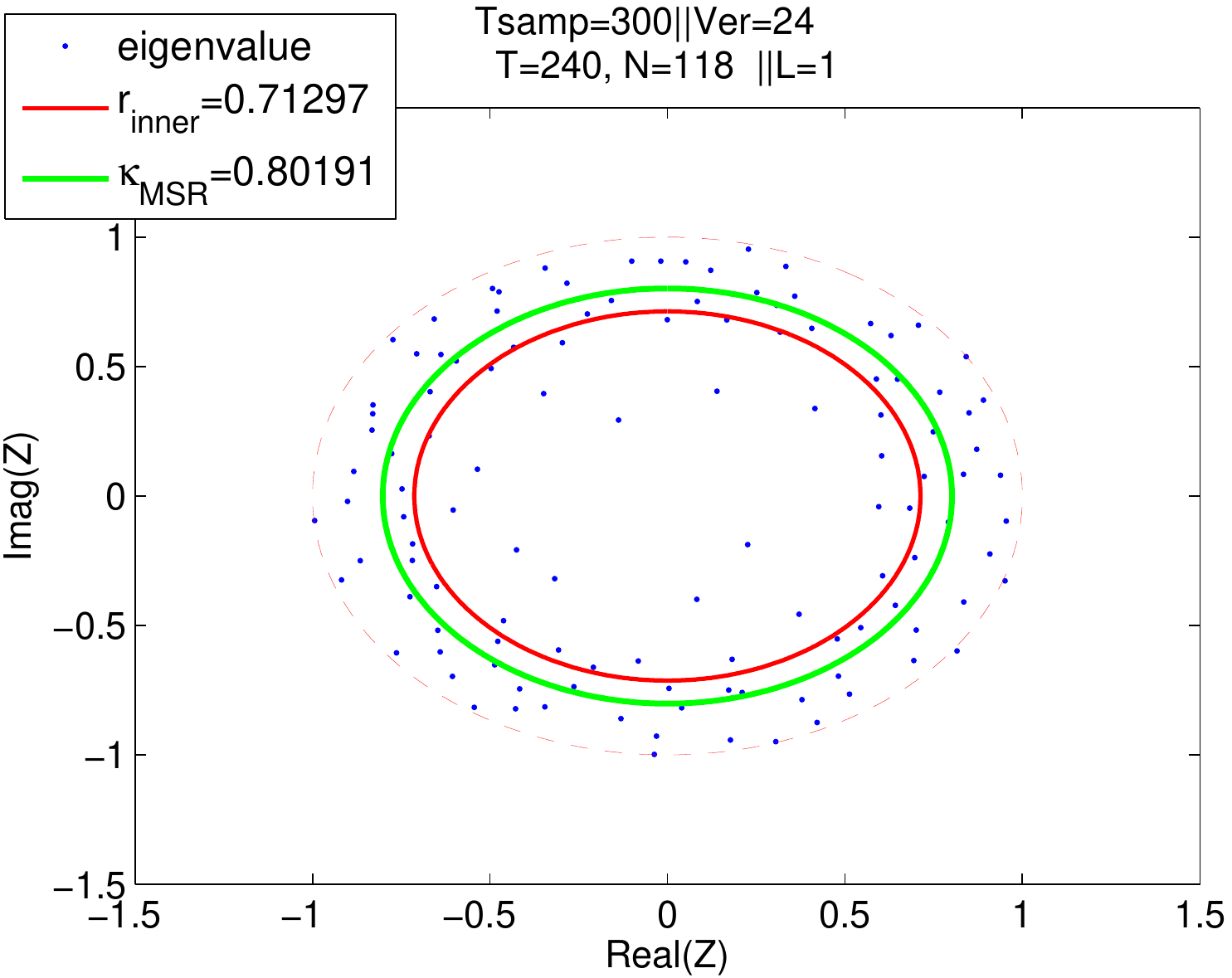}{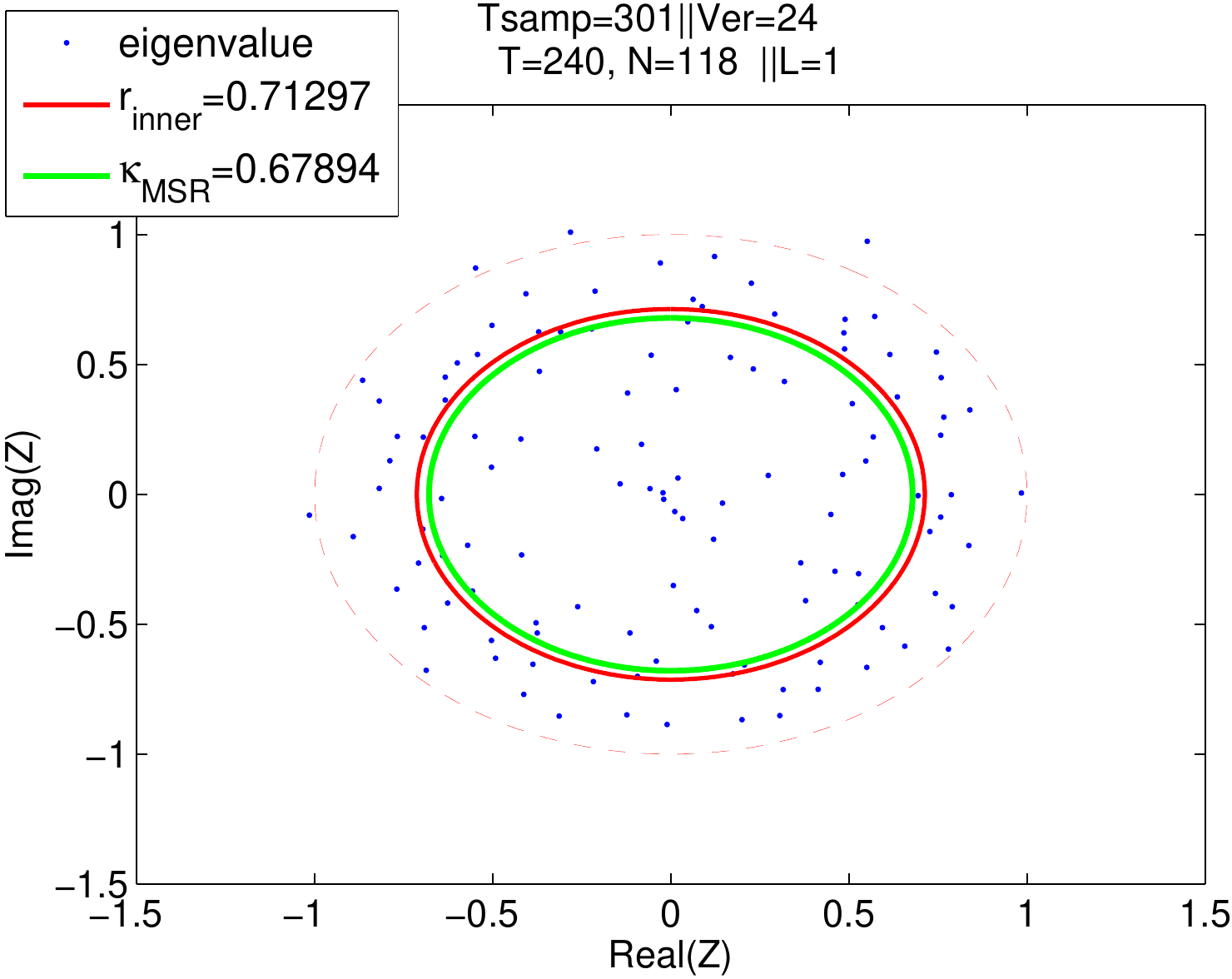}{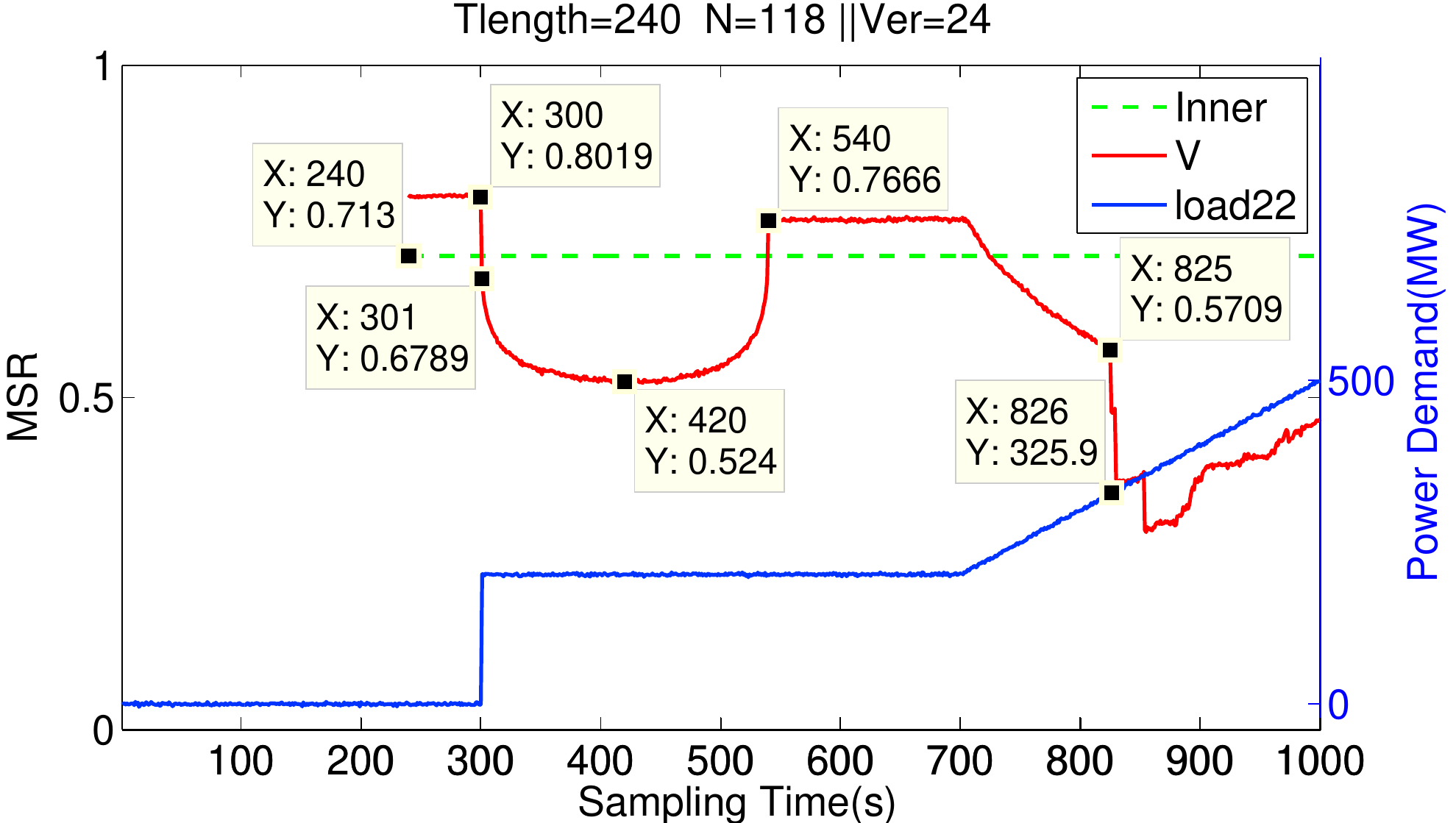}
{Ring Law and MSR at Different Sampling Times}
{htbp} 

At sampling time \VtSS{300}{s}, the data include a time area \VATT {61\!:\!300}{s}. The noise plays a dominant part, and the distribution of eigenvalues \Eig Z is more closely to the standard ring of \Data {L=1}. At sampling time \VtSS{301}{s}, \ref{fig:Case Ringb} shows that the eigenvalues collapse to the center point of the circle.

From \Cur {\VMSR}{t} curve in Figure \ref{fig:Case Ringc}, it is observed that the \VMSR{} starts to decrease  {\Data {\small(0.8019, 0.6789,\cdots,0.5240)}} just at \Vtt{301}{s} when the event of sudden change occurs as signal. This decreasing lasts for half of the time length \mbox{(i.e. \Equs{T/2}{120} s)}. The statistic \VMSR{} is a high-dimensional statistic for plotting \Cur {\VMSR}{t} curve which is sensitive to the event. The \Cur {\VMSR}{t} curve of each partition can be plotted in the same way. Both the voltage \VV {} and the high-dimensional statistic \VMSR{} are able to be achieved from the raw data \Rdata {\VV {}} by a statistical procedure without any engineering model. The former is oriented to a single point, whose value is fully decided by a raw data at a certain single sampling time; and the latter is related to a high-dimensional matrix, whose value is decided by all matrix entries consisting raw data of \Data {N} varieties during \Data {T} sampling times.

Based on the statistical analysis above, we conduct our visualization. It will come to a conclusion that the combination of high-dimensional analysis and visualization has a better performance in watching the status and trends of the power system.

\begin{figure*}[!t]
\centering
\subfloat[\VtSS{300}{s}]{
\includegraphics[width=0.16\textwidth]{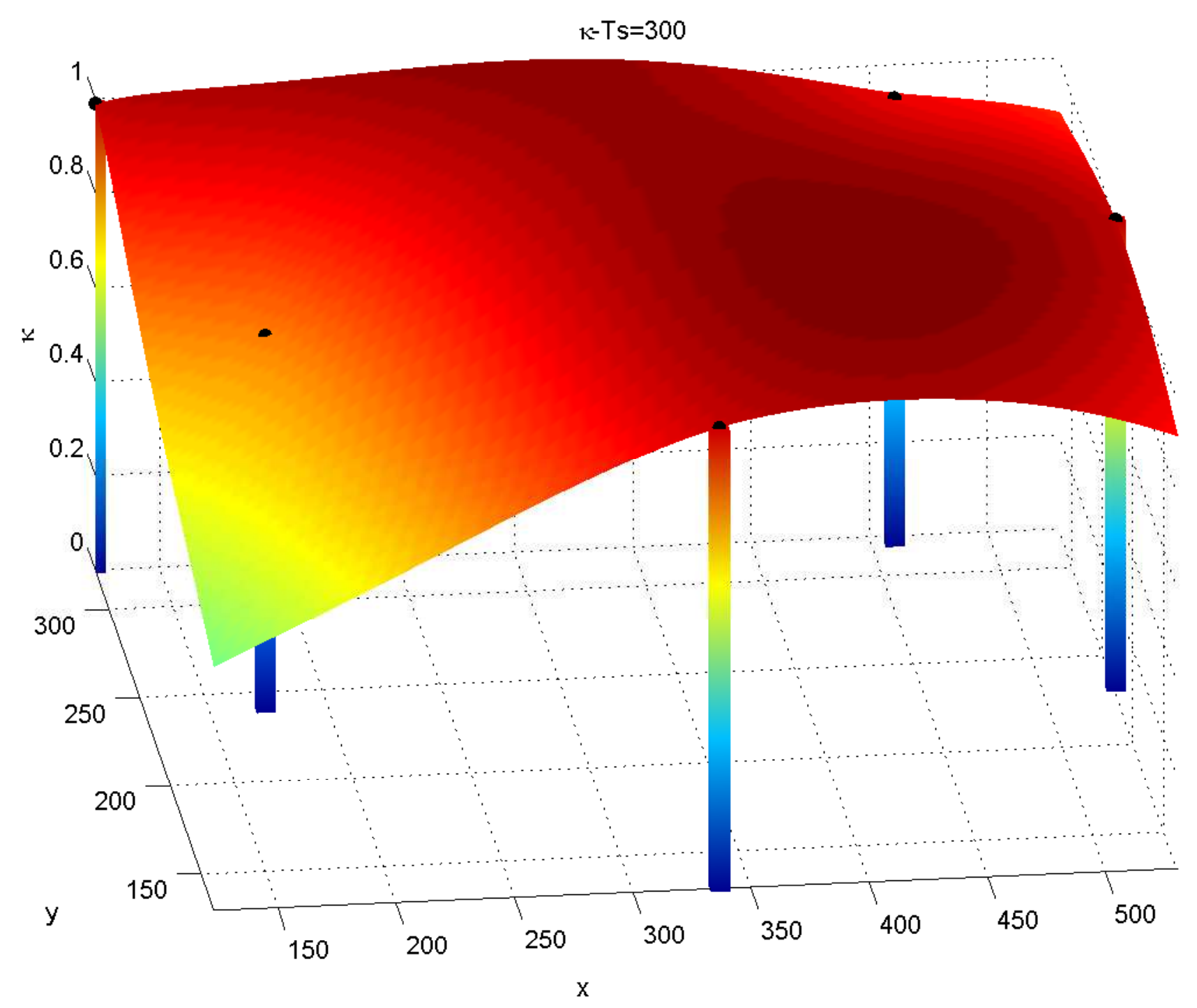}
}
\subfloat[\VtSS{301}{s}]{
\includegraphics[width=0.16\textwidth]{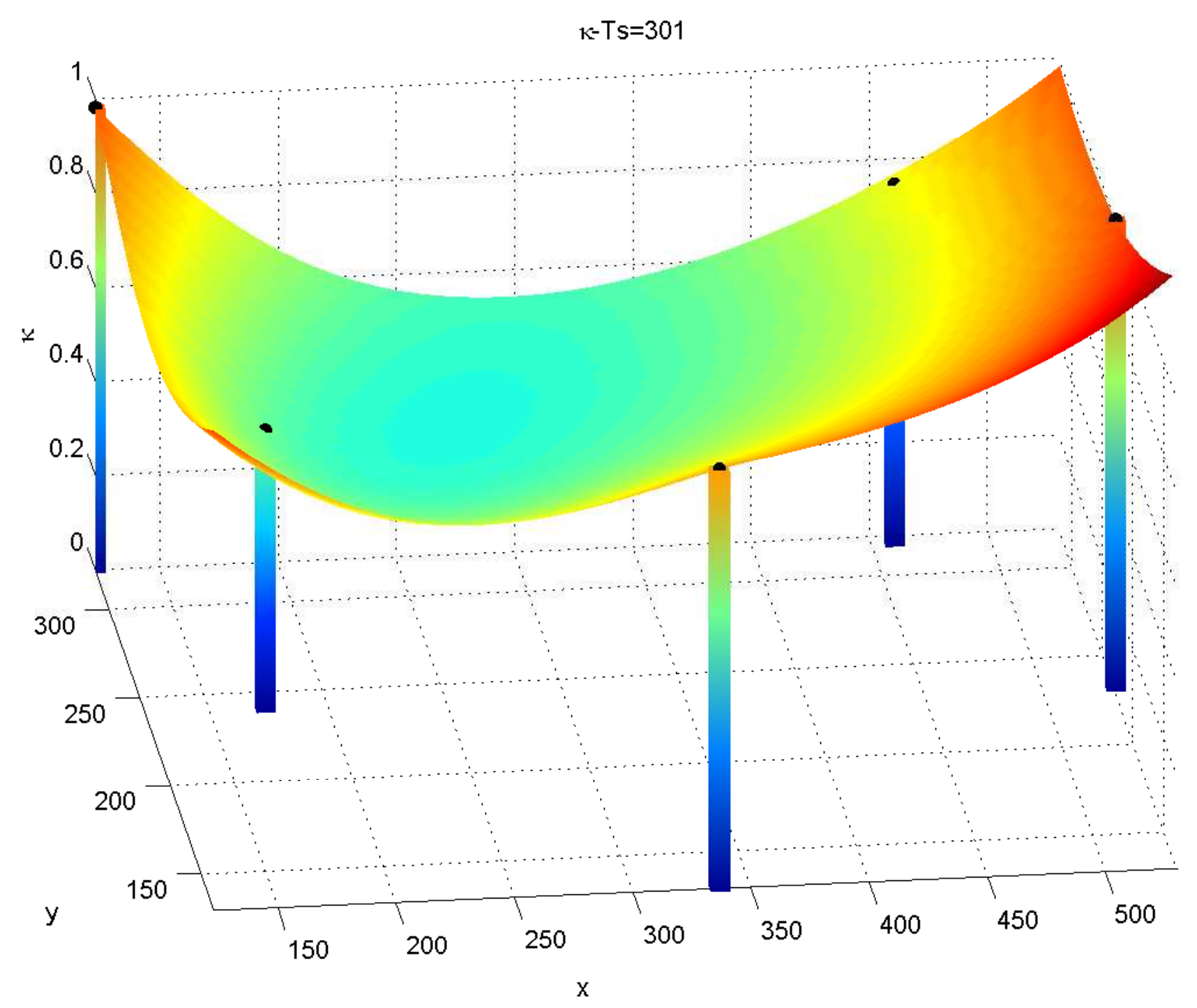}
}
\subfloat[\VtSS{302}{s}]{
\includegraphics[width=0.16\textwidth]{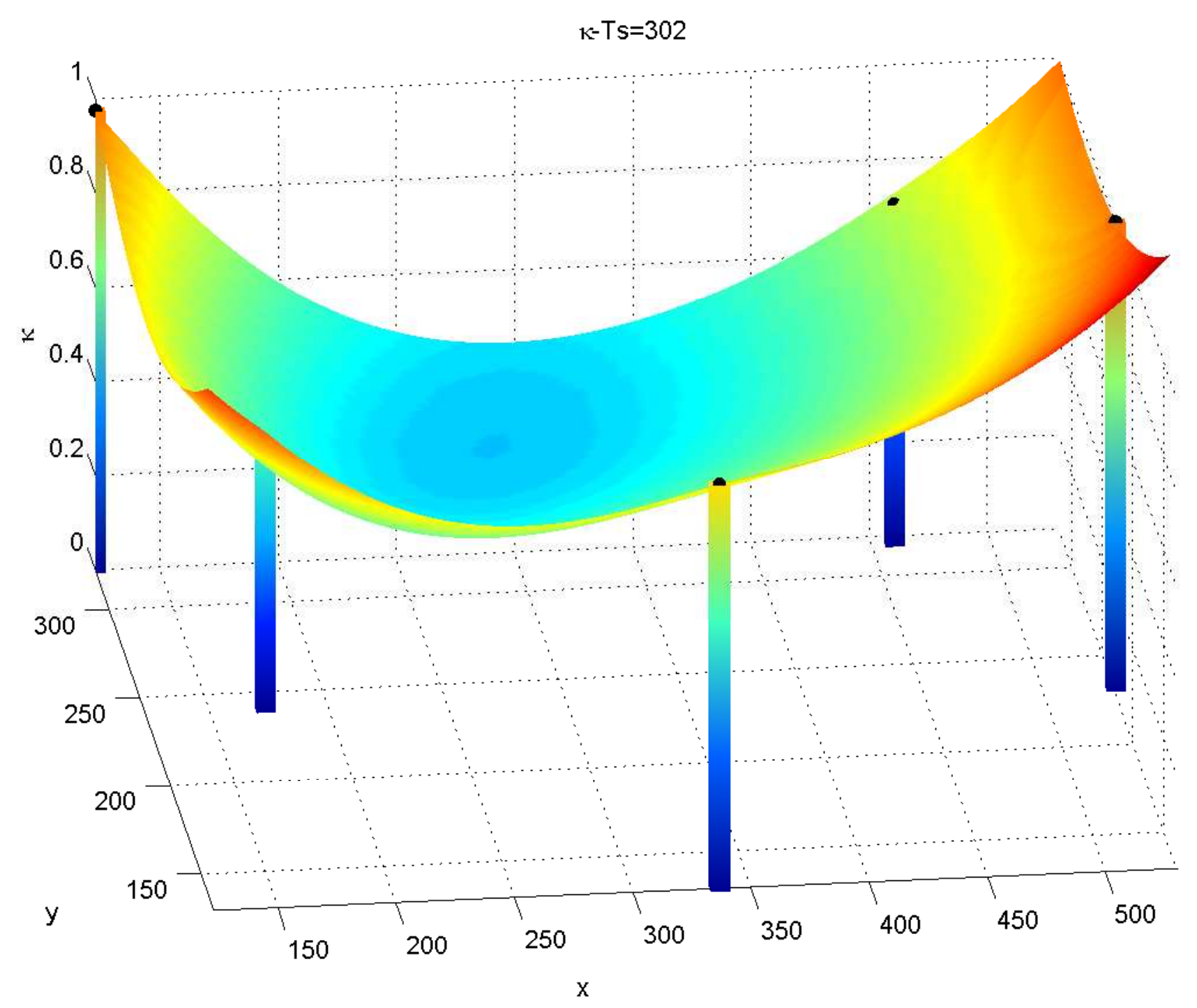}
}
\subfloat[\VtSS{420}{s}]{
\includegraphics[width=0.16\textwidth]{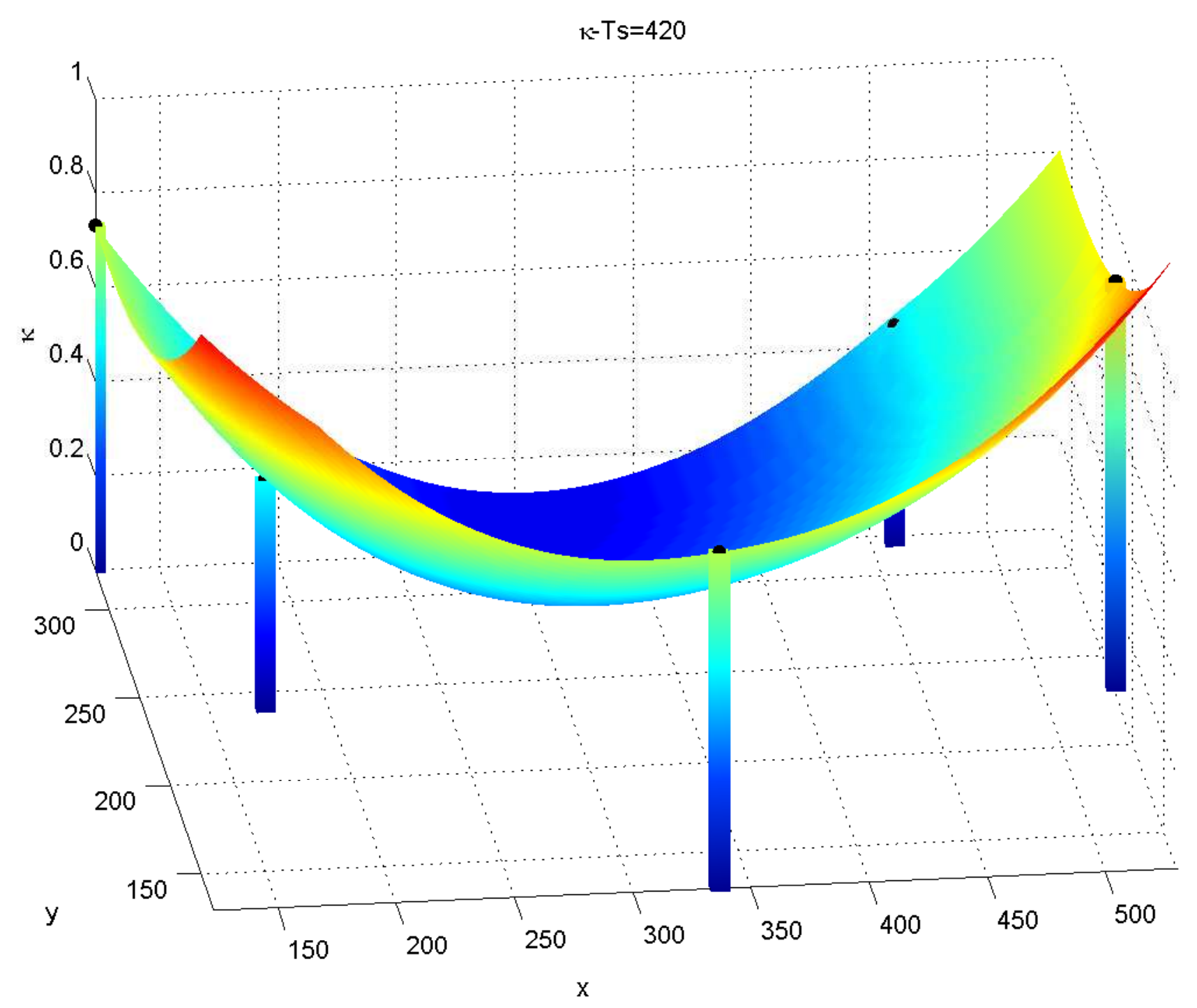}
}
\subfloat[\VtSS{820}{s}]{
\includegraphics[width=0.16\textwidth]{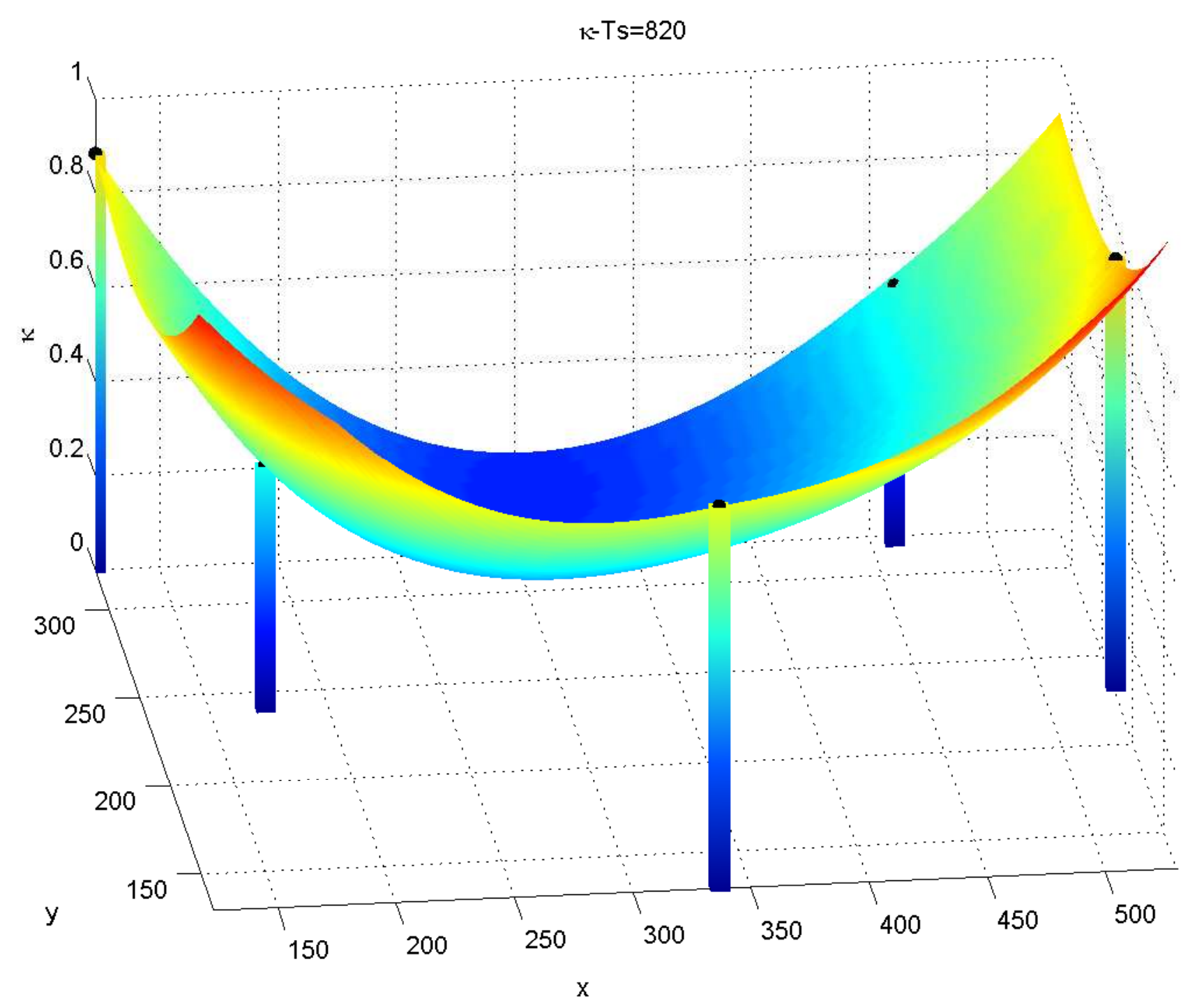}
}
\subfloat[\VtSS{826}{s}]{
\includegraphics[width=0.16\textwidth]{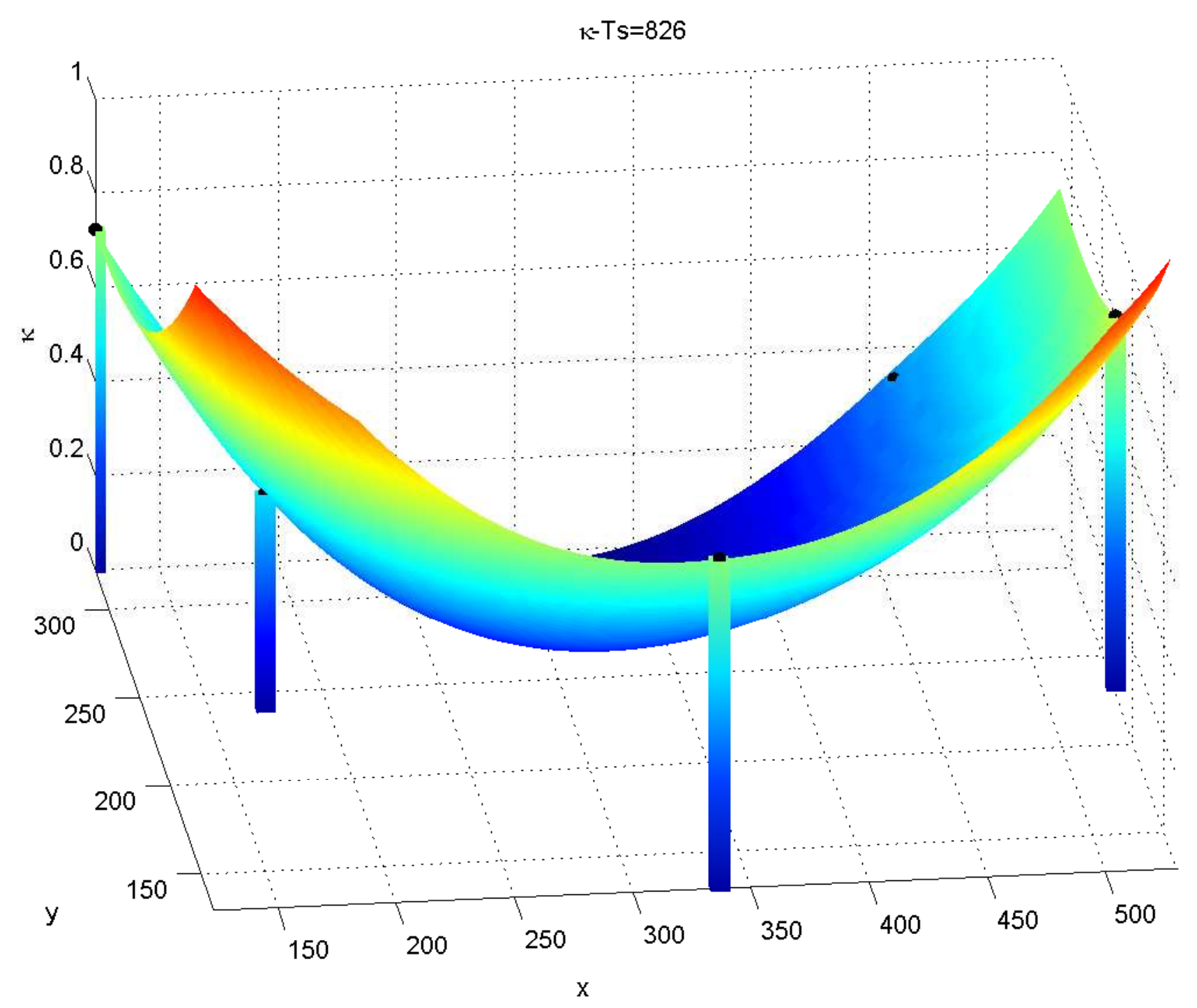}
}

\caption{Visualization of the High-dimensional Index \VMSR{} with Full Data Sets}
\label{fig:FullDataMSR}

\centering
\subfloat[\VtSS{300}{s}]{
\includegraphics[width=0.16\textwidth]{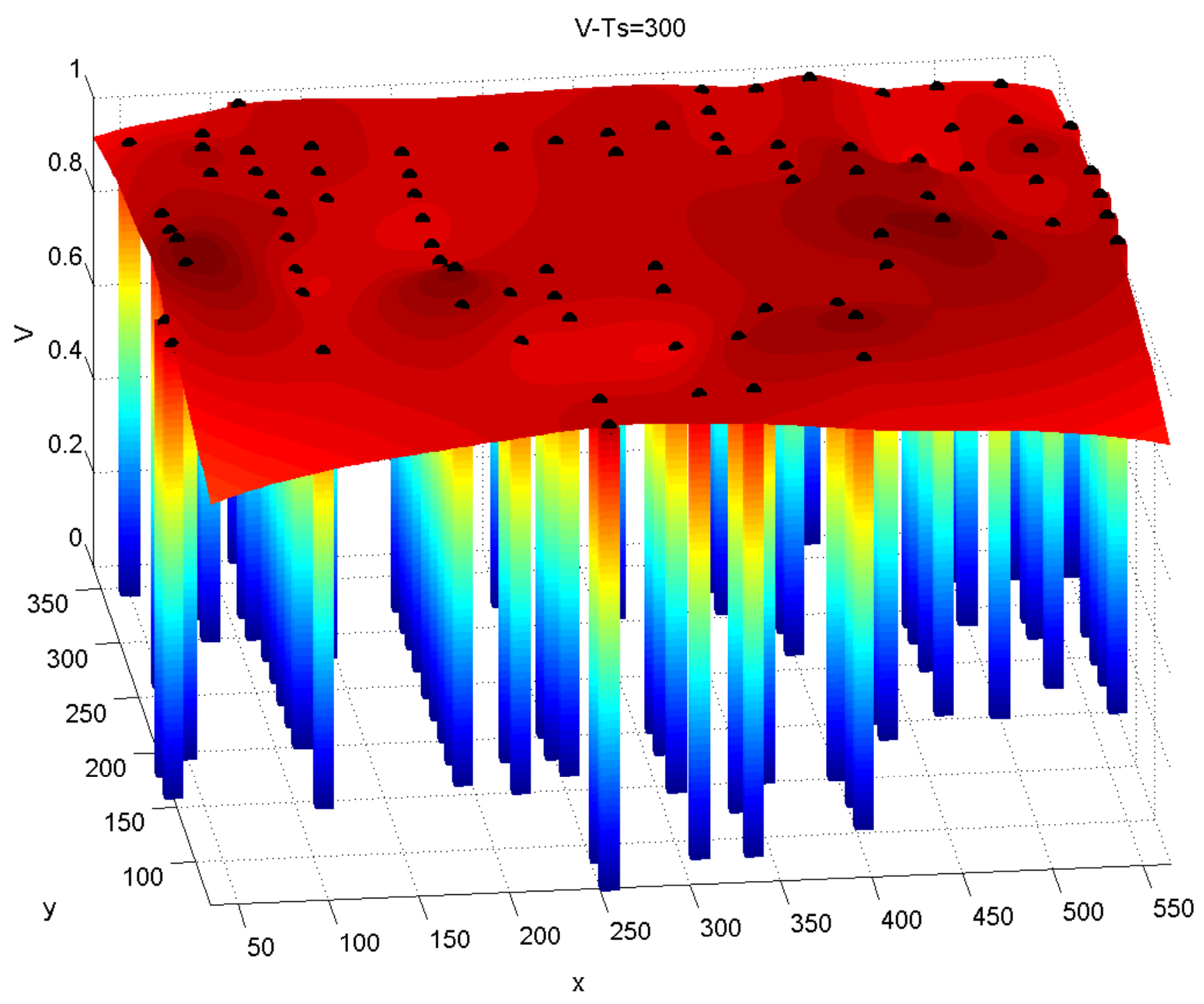}
}
\subfloat[\VtSS{301}{s}]{
\includegraphics[width=0.16\textwidth]{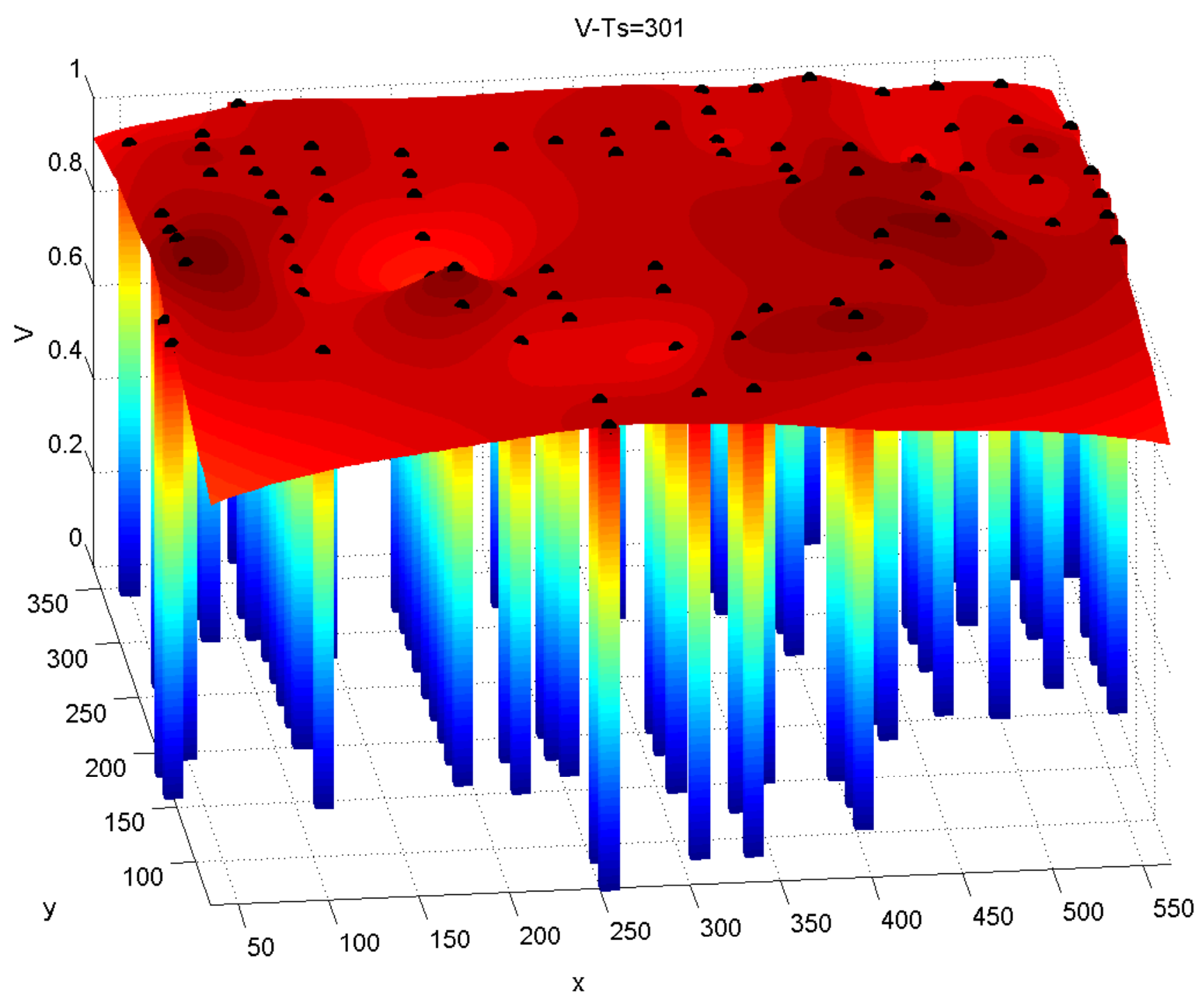}
}
\subfloat[\VtSS{302}{s}]{
\includegraphics[width=0.16\textwidth]{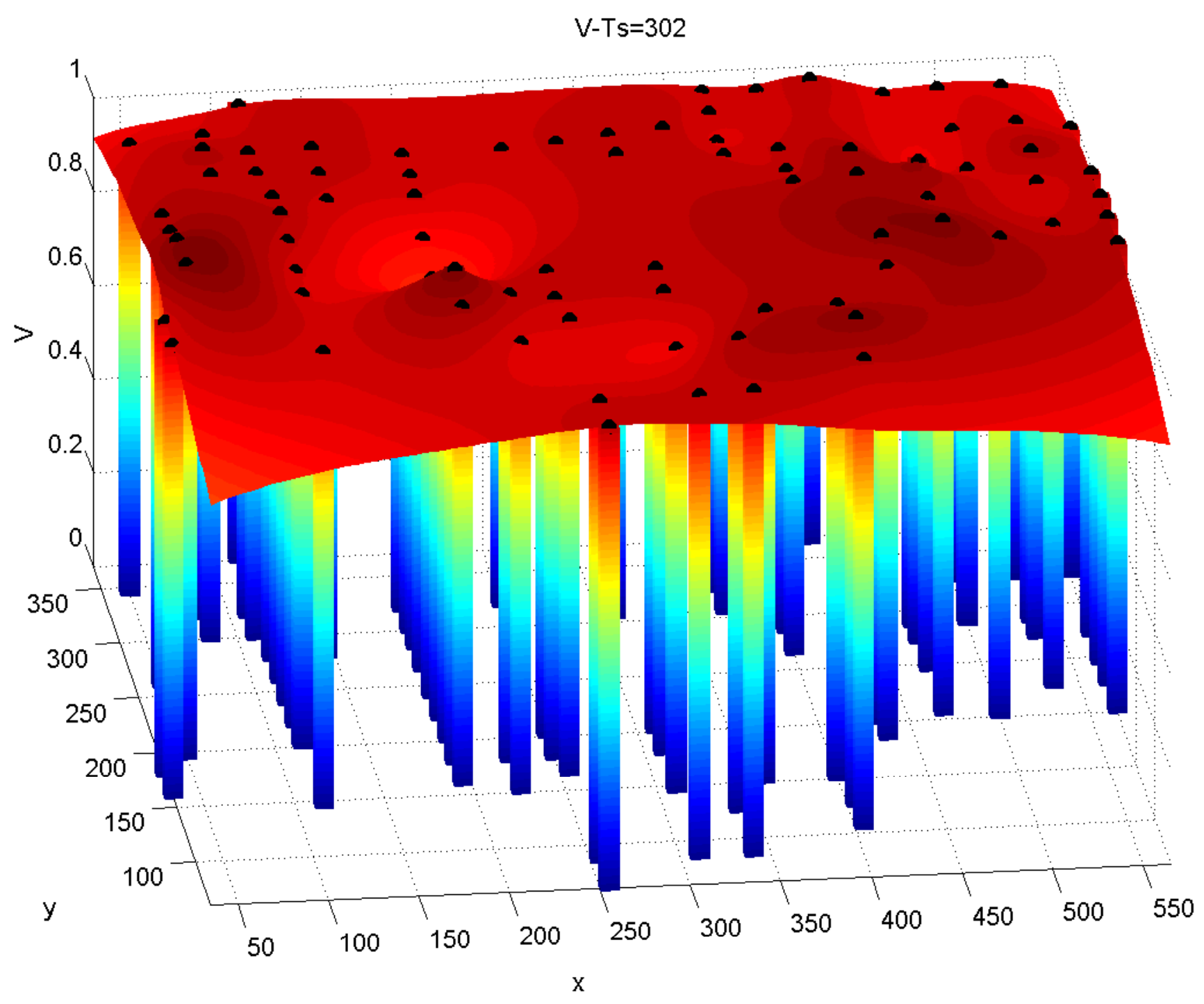}
}
\subfloat[\VtSS{420}{s}]{
\includegraphics[width=0.16\textwidth]{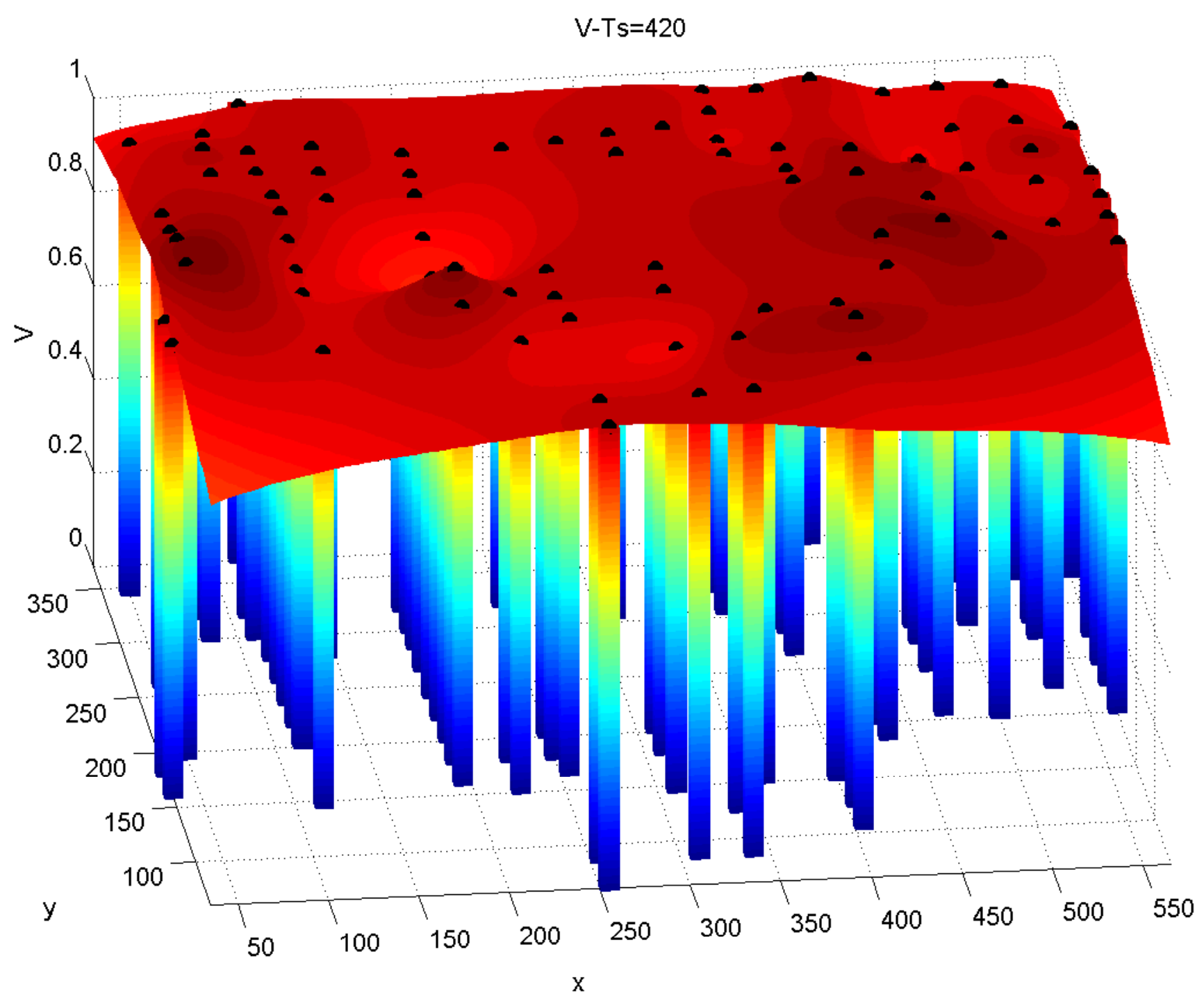}
}
\subfloat[\VtSS{820}{s}]{
\includegraphics[width=0.16\textwidth]{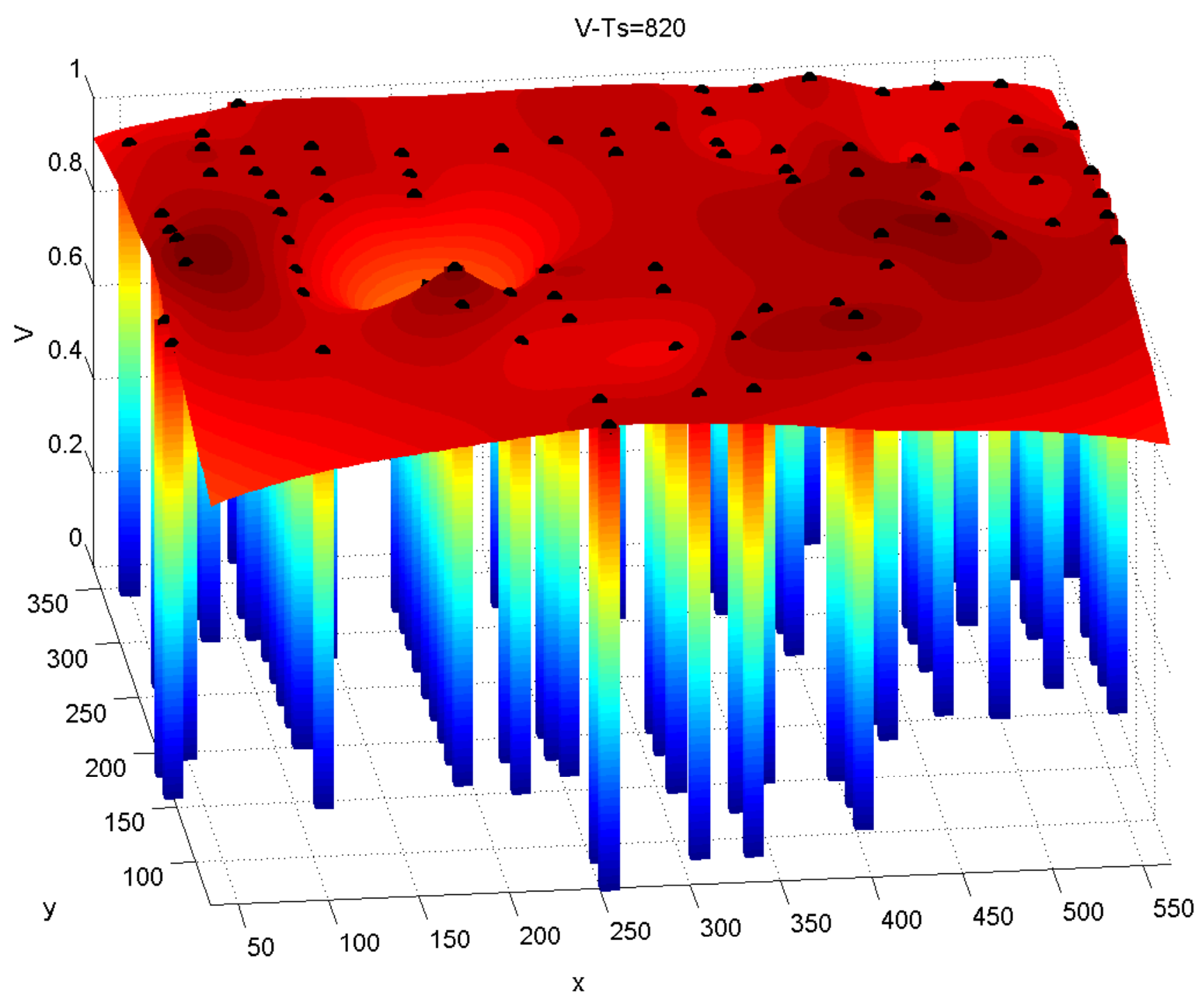}
}
\subfloat[\VtSS{826}{s}]{
\includegraphics[width=0.16\textwidth]{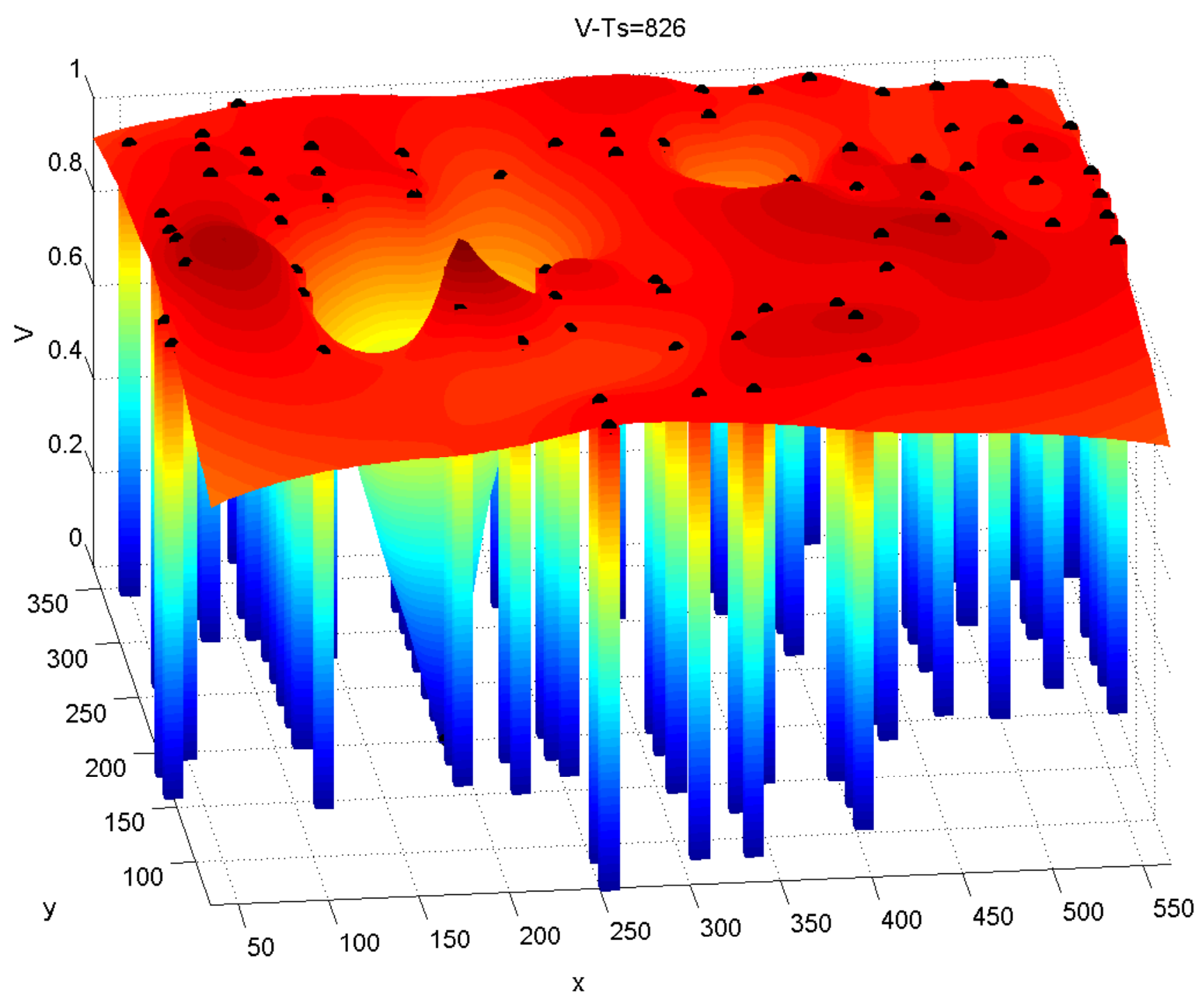}
}

\caption{Visualization of the Voltage \VV{} with Full Data Sets}
\label{fig:FullDataV}

\centering
\subfloat[\VtSS{300}{s}]{
\includegraphics[width=0.16\textwidth]{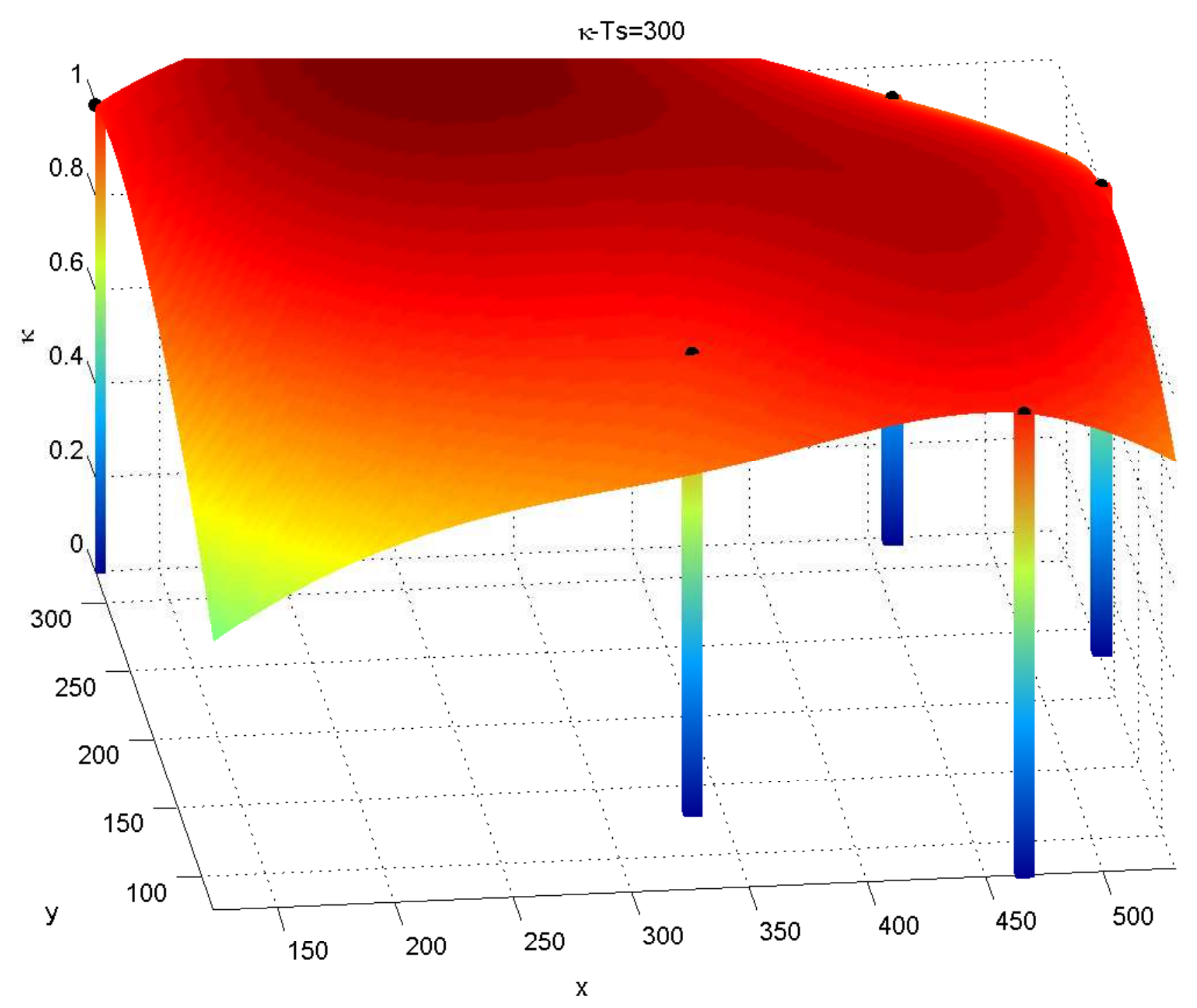}
}
\subfloat[\VtSS{301}{s}]{
\includegraphics[width=0.16\textwidth]{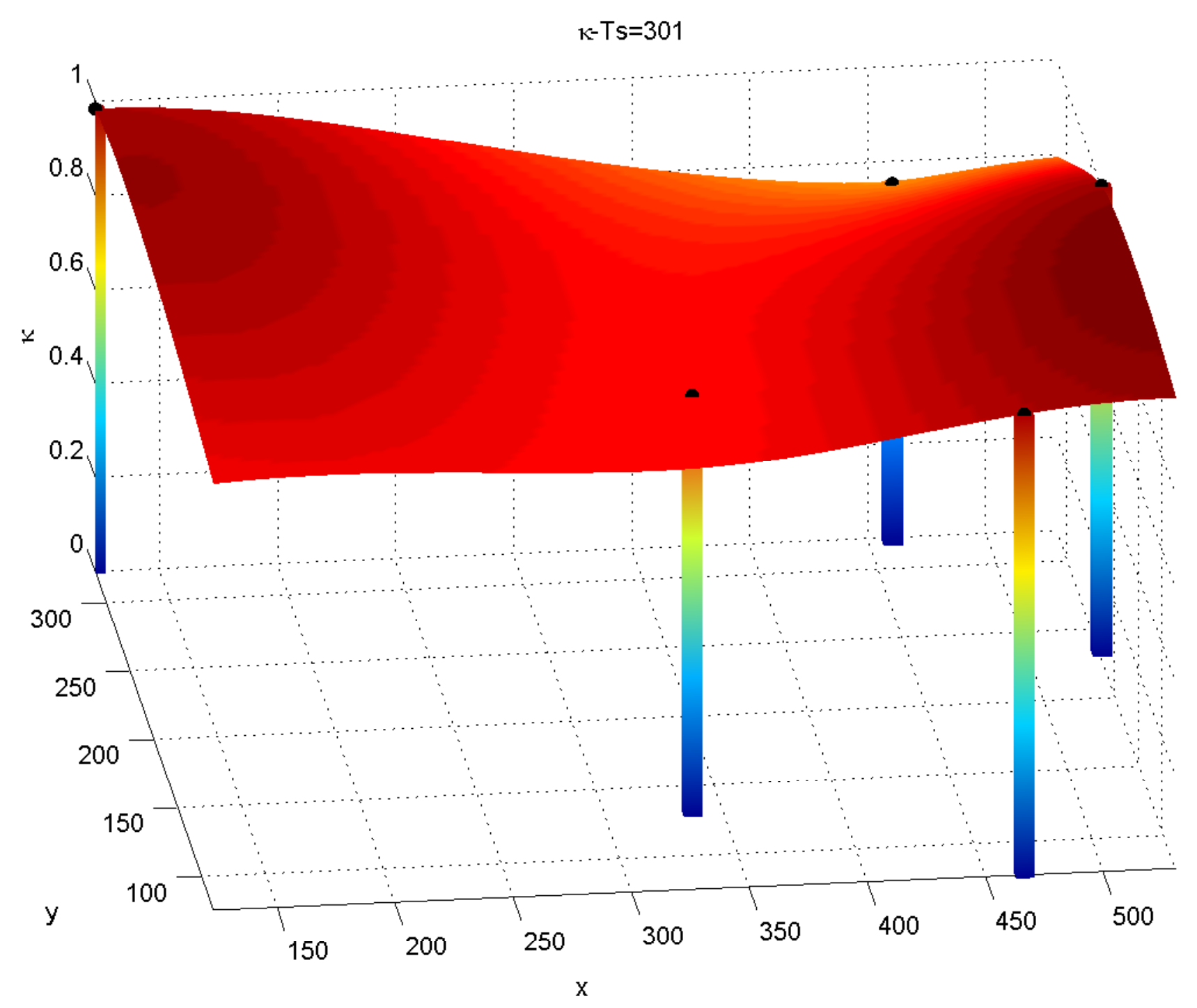}
}
\subfloat[\VtSS{302}{s}]{
\includegraphics[width=0.16\textwidth]{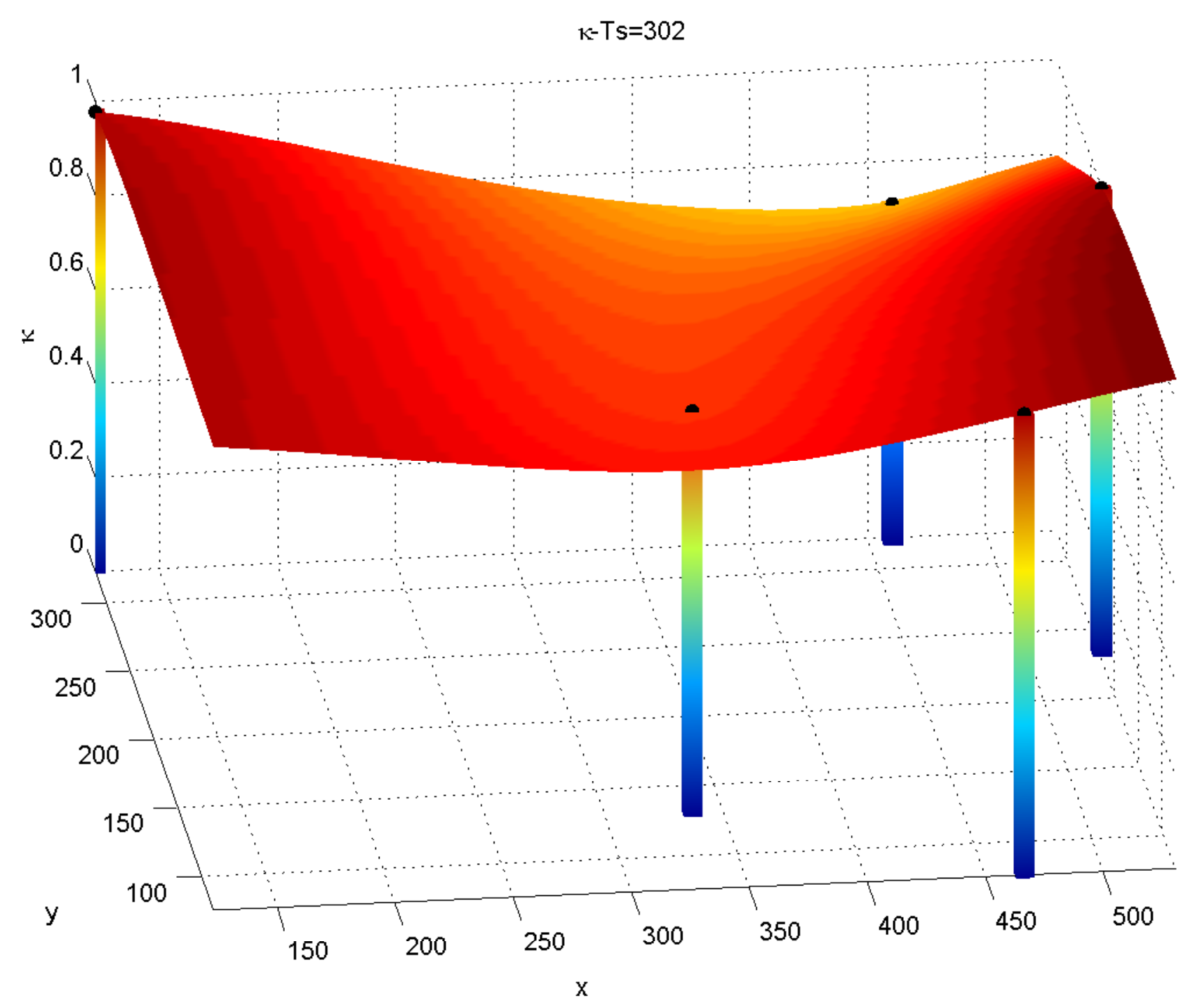}
}
\subfloat[\VtSS{420}{s}]{
\includegraphics[width=0.16\textwidth]{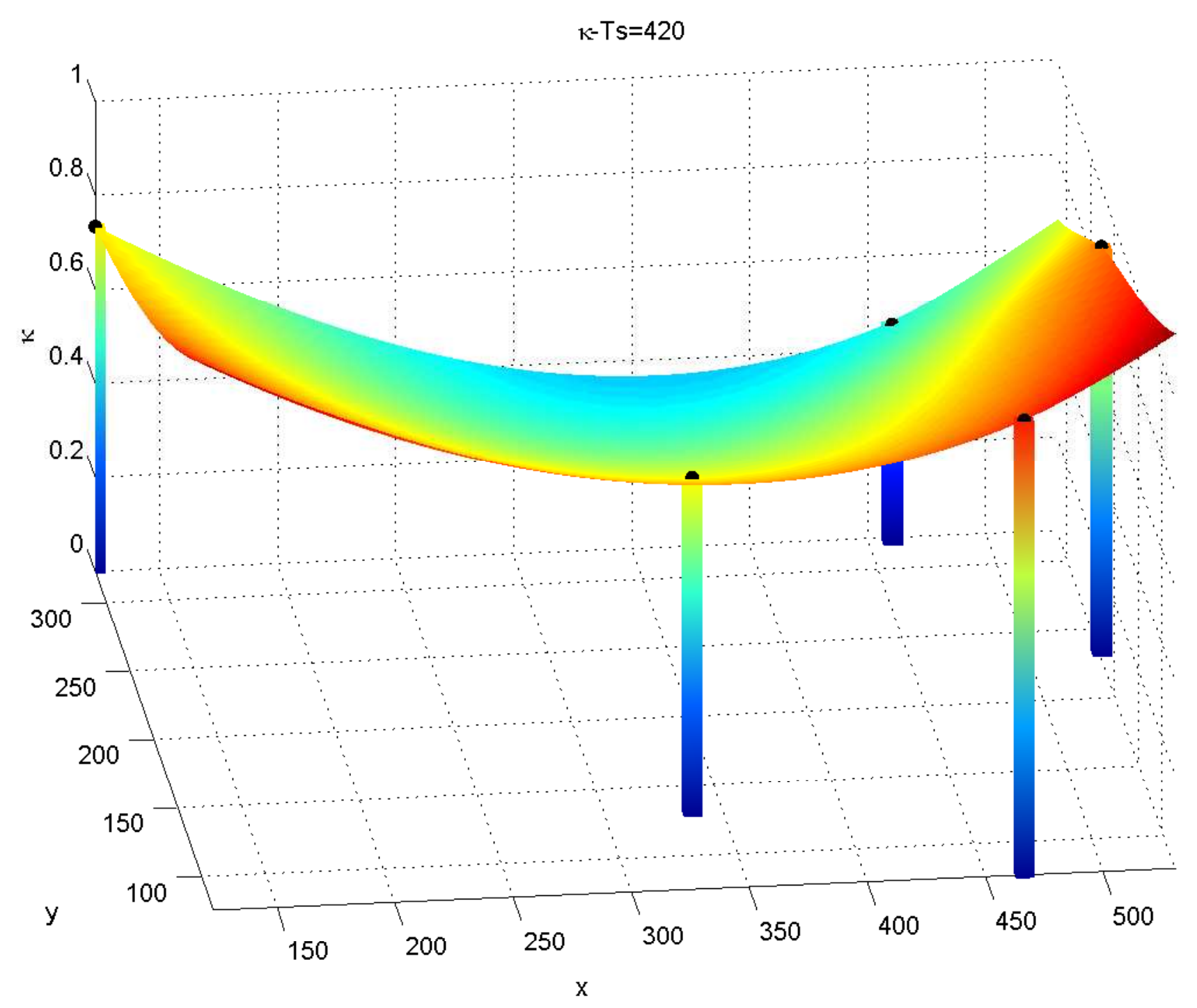}
}
\subfloat[\VtSS{820}{s}]{
\includegraphics[width=0.16\textwidth]{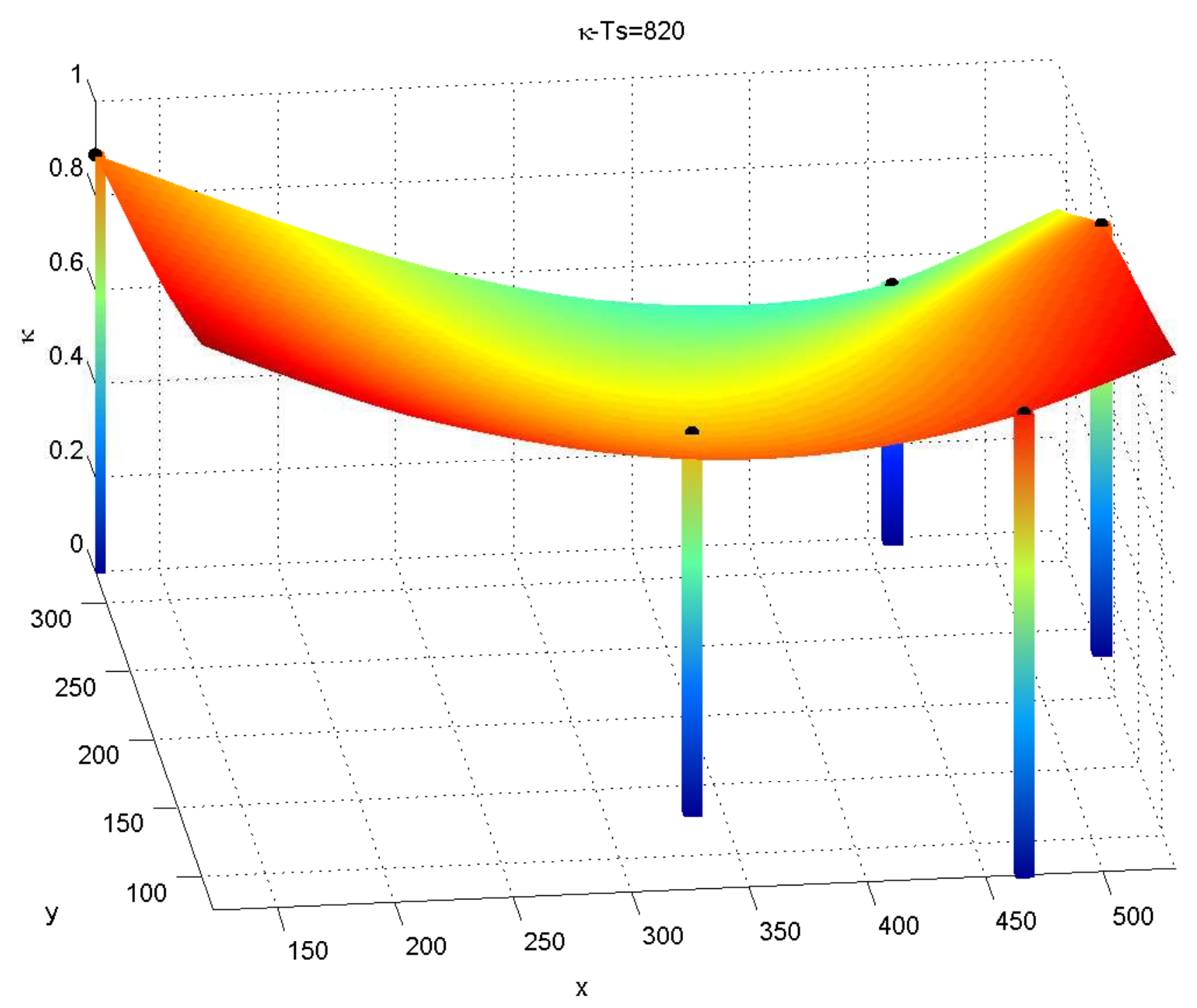}
}
\subfloat[\VtSS{826}{s}]{
\includegraphics[width=0.16\textwidth]{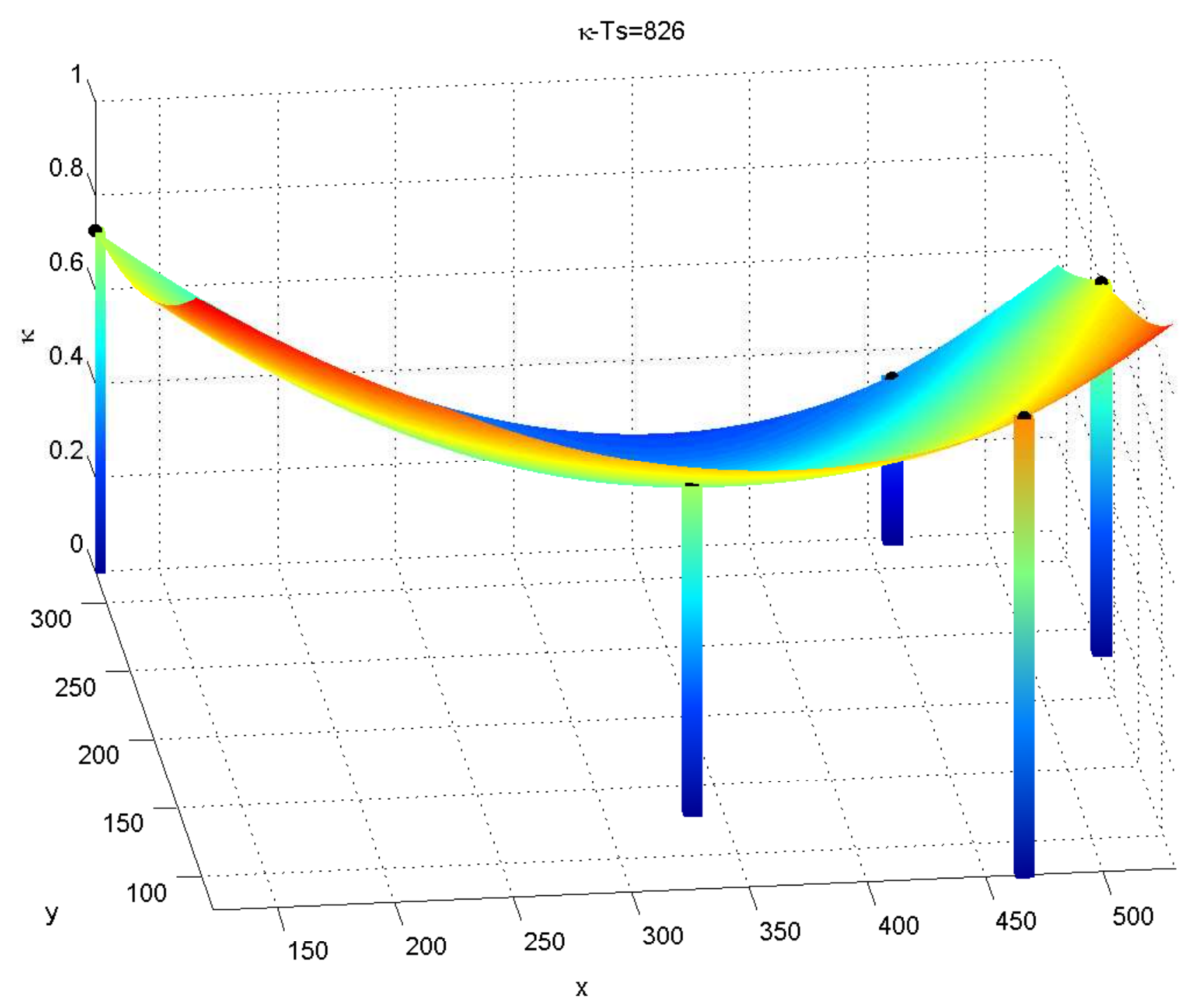}
}

\caption{Visualization of the High-dimensional Index \VMSR{} without Data Sets of A2}
\label{fig:DataWithoutA2MSR}

\centering
\subfloat[\VtSS{300}{s}]{
\includegraphics[width=0.16\textwidth]{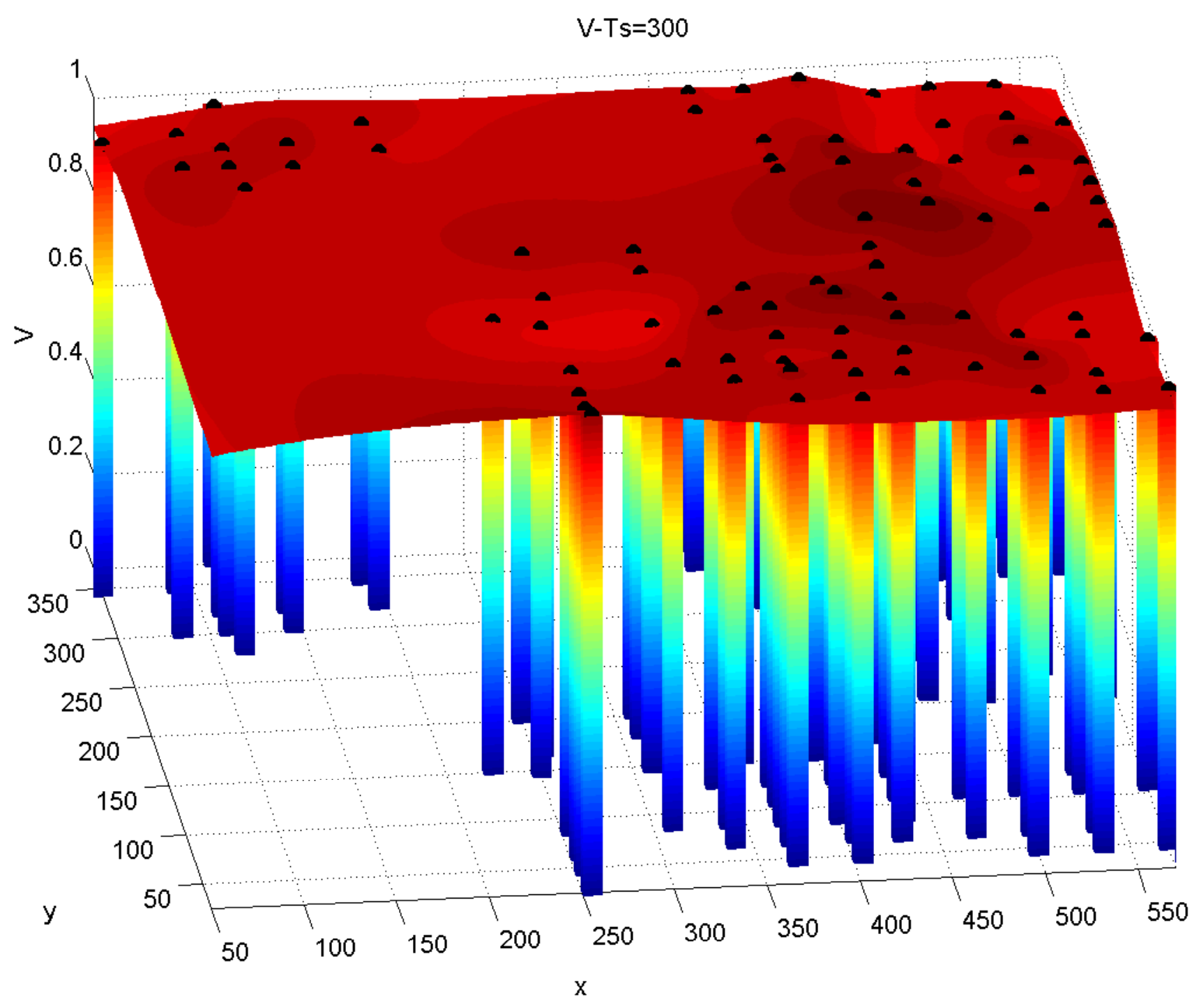}
}
\subfloat[\VtSS{301}{s}]{
\includegraphics[width=0.16\textwidth]{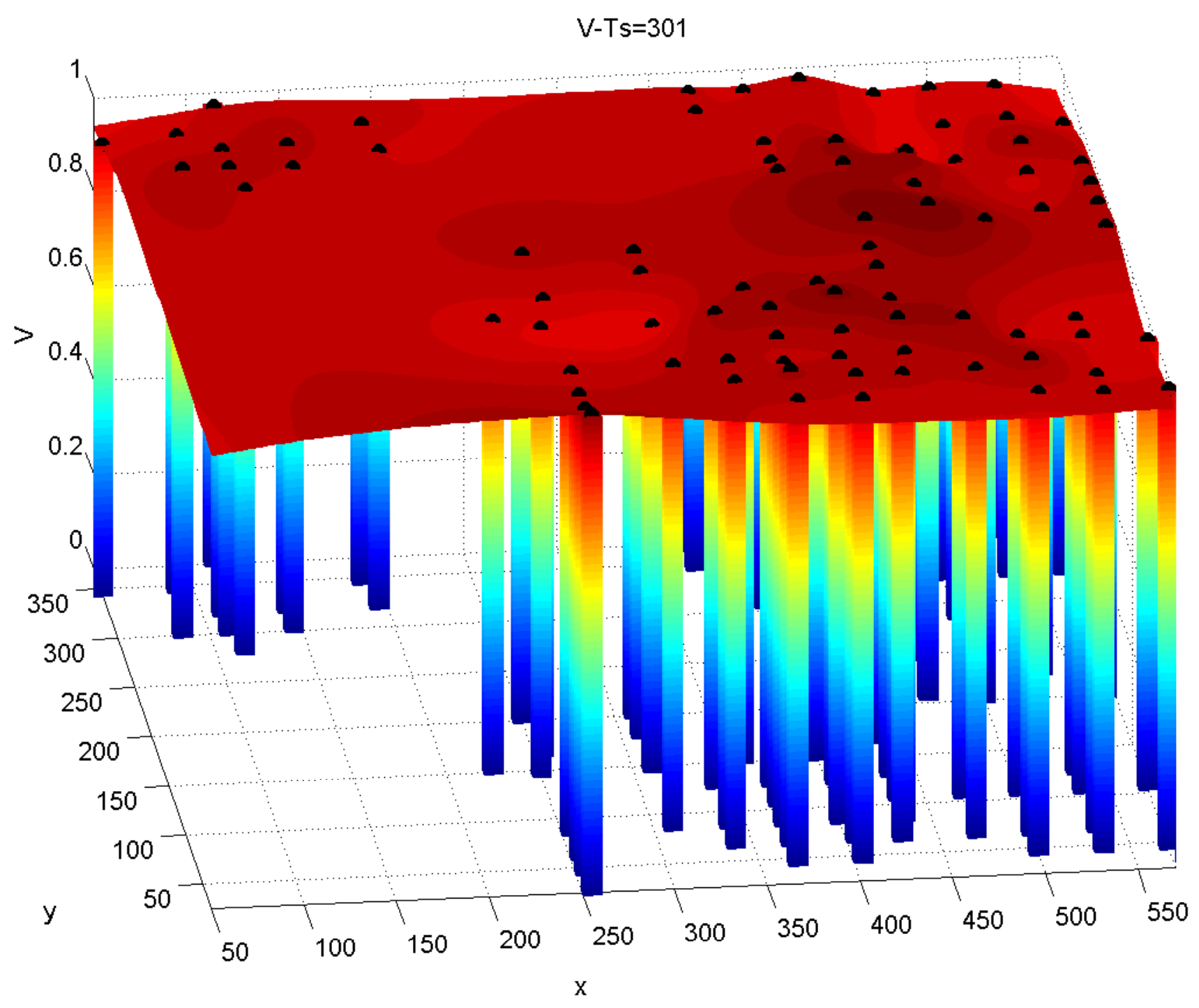}
}
\subfloat[\VtSS{302}{s}]{
\includegraphics[width=0.16\textwidth]{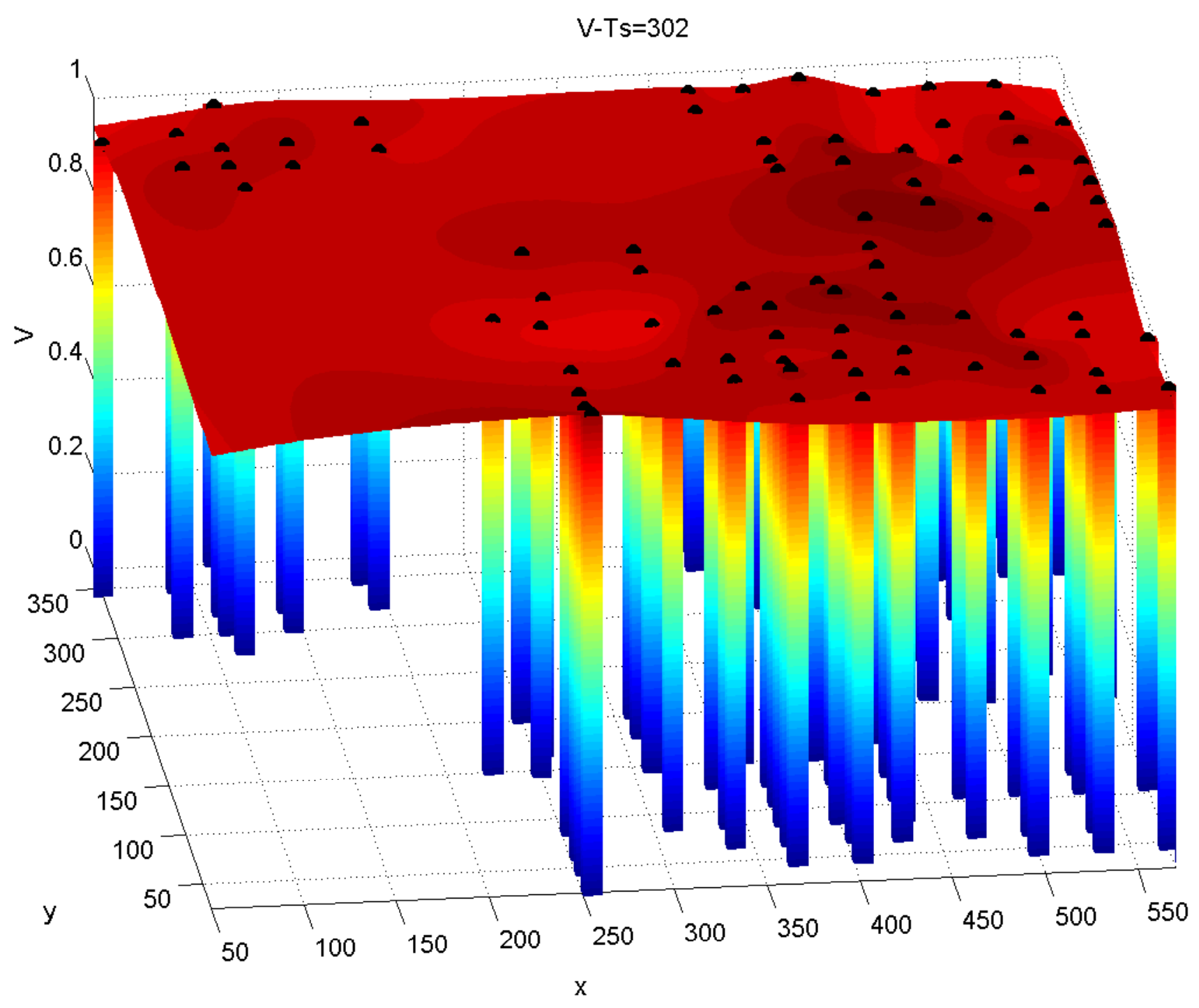}
}
\subfloat[\VtSS{420}{s}]{
\includegraphics[width=0.16\textwidth]{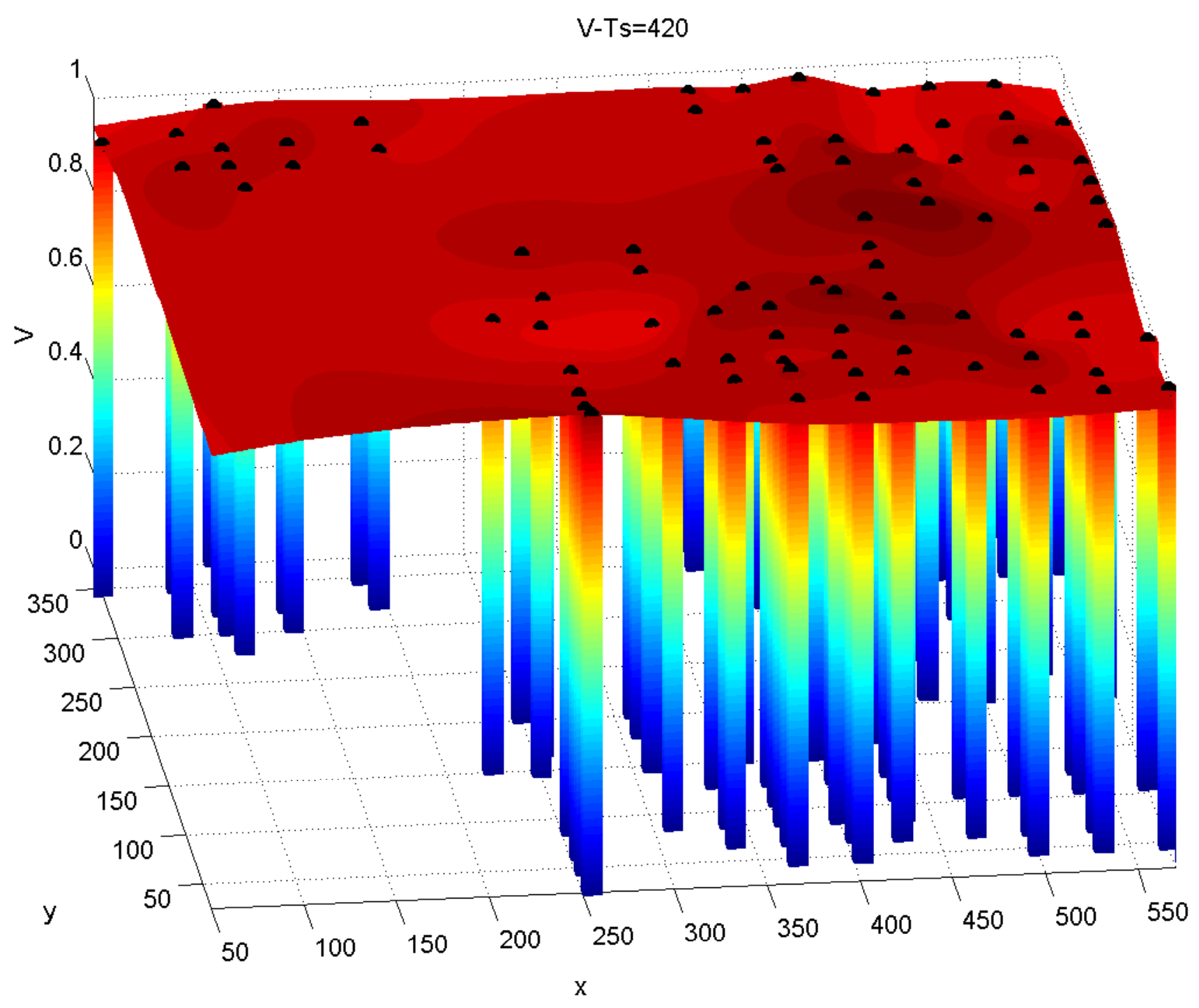}
}
\subfloat[\VtSS{820}{s}]{
\includegraphics[width=0.16\textwidth]{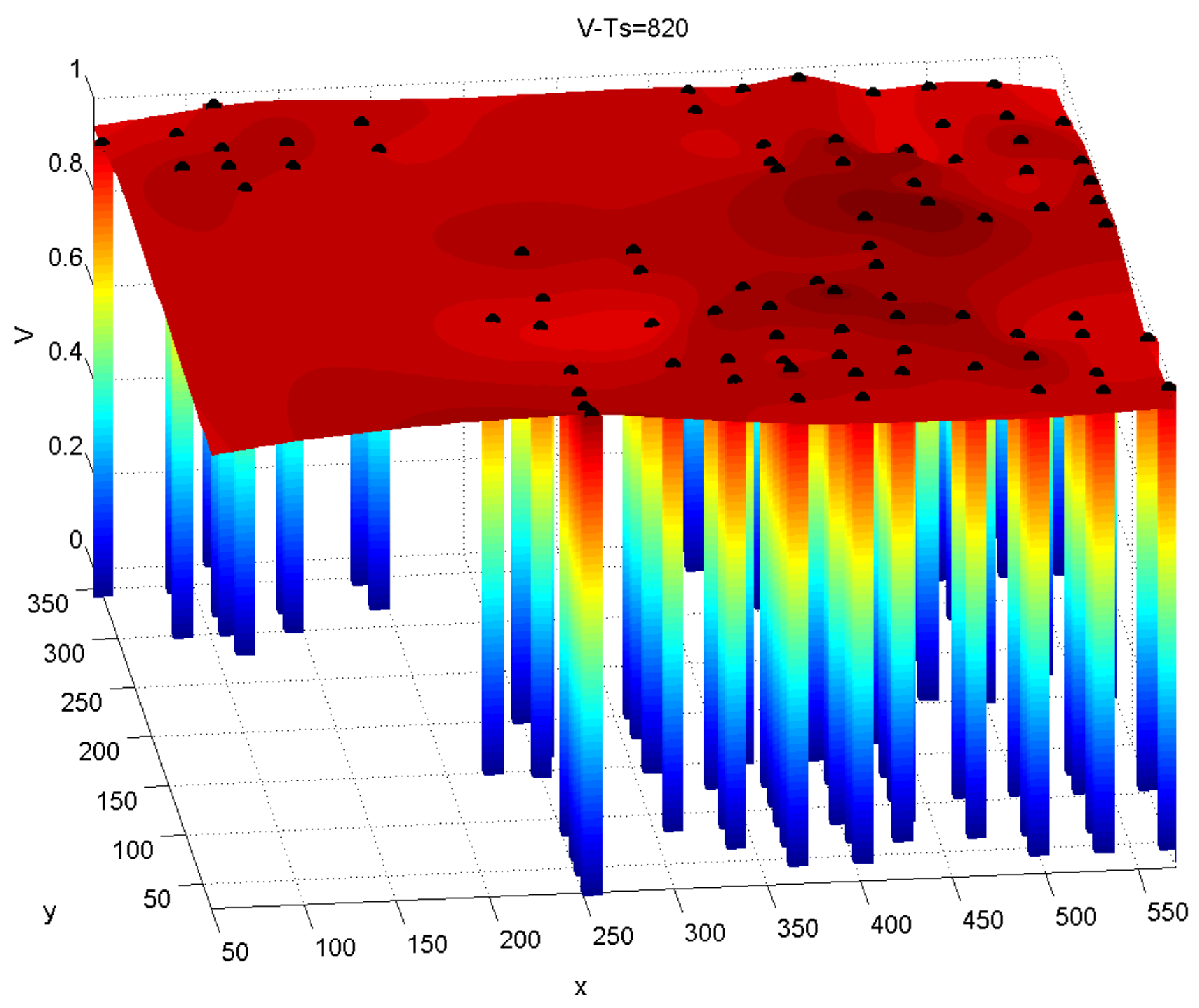}
}
\subfloat[\VtSS{826}{s}]{
\includegraphics[width=0.16\textwidth]{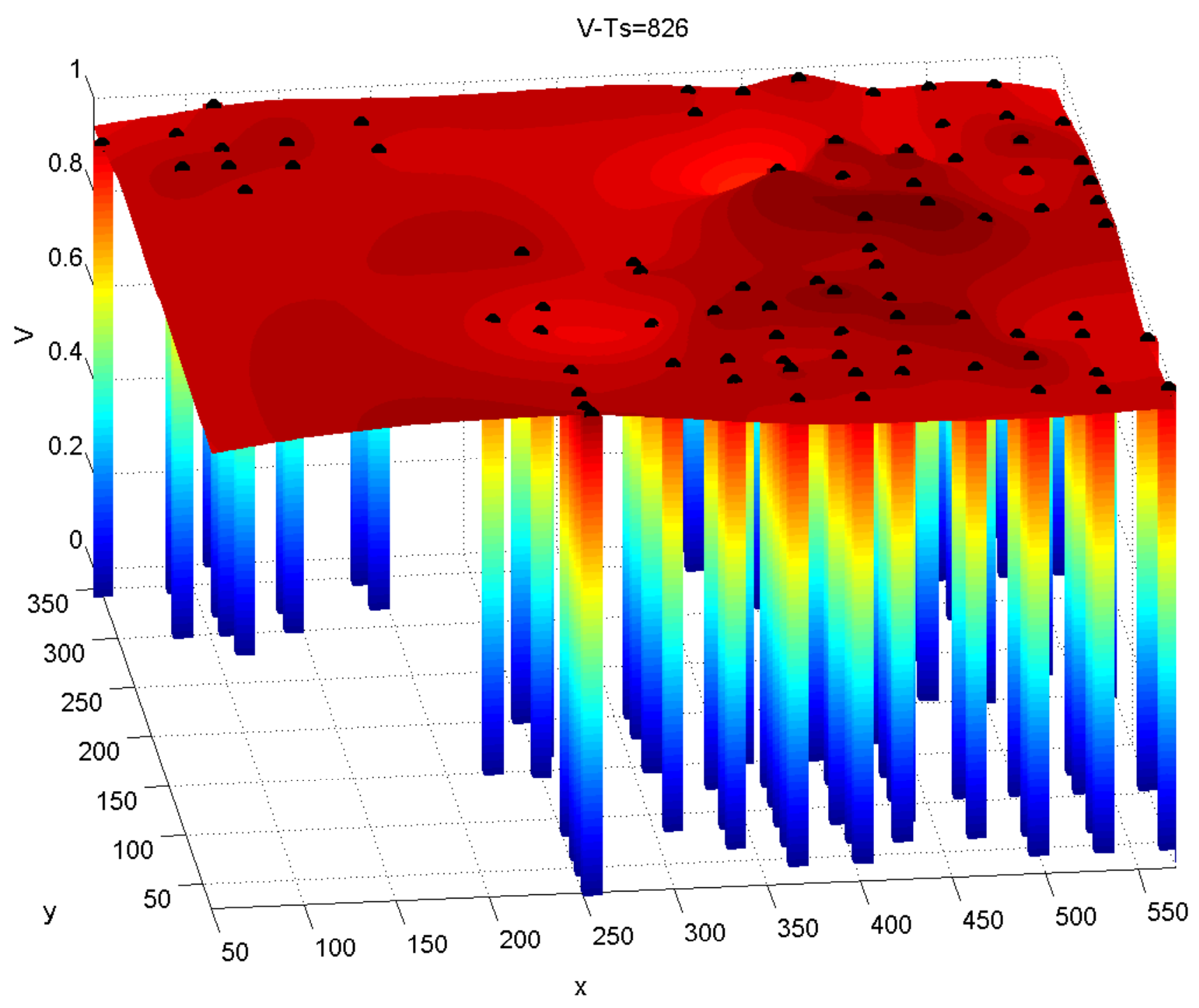}
}

\caption{Visualization of the Voltage \VV{} without Data Sets of A2}
\label{fig:DataWithoutA2V}
\end{figure*}

With an interpolation method \cite{weber2000voltage}, a 3D power-map consisting of data is plotted. Figure \ref{fig:FullDataMSR} and \ref{fig:FullDataV} depict some key frame in 3D power-map animation of the high-dimensional statistic \VMSR{} and of the raw data \Rdata {\VV {}}, respectively. Comparing Figure \ref{fig:FullDataMSR} and \ref{fig:FullDataV}, it is concluded that the high-dimensional statistic \VMSR{} is sensitive to the events and performs better in watching the status and trend:

a) At time \Vtt{301}{s}, area around A2 in the power-map changes. Therefore, we conjecture some events occur at A2. The performance during times \Vtt{302\!:\!420}{s} validates this conjecture. Moreover, we can conjecture that the events are more influential in the areas of A2, A3, A5 and have no effect on A6; and the conjecture coincides with the reality that there is a sudden change of \VPbus{22} at \Vtt{301}{s}.

b) With sustainable growth of power demand at some bus (\VPbus{n}), the whole system becomes more and more vulnerable. The vulnerability can be estimated by the visualization of \VMSR{}.

c) Moreover, if the most important data (i.e. raw data \Rdata {\VV {}} for A2) is lost somehow, hardly any valuable information can be got by \VV {} as Figure \ref{fig:DataWithoutA2V}, whereas the status and trend of the whole system can still be estimated by \VMSR{} as Figure \ref{fig:DataWithoutA2MSR}.

d) At last, we conjecture that the low-dimensional statistic, such as raw data at a single time, 1-dimensional statistic mean \Data \mu{}  and 2-dimensional statistic variance \Data \sigma, cannot reflect the status and trend as well as the high-dimensional statistic \VMSR{}.

%

%



\section{Conclusion}
This paper develops a 3D power-map animation by integrating high-dimensional analysis and visualization. For the former, we introduced the single-ring law and statistic MSR; and for the latter, we utilized distributed MSR data and interpolation method. Case study validated the effectiveness and performance of the 3D map in watching the status and trend of the whole system. Especially, its robustness against the bad data is a highlight---the 3D map of high-dimensional data is able to conduct estimation for power systems even with loss of the data in the most related partition.

However, there are still some questions left. For example, to figure out the relationships between the mean spectral energy radius MSR and the physical parameters is a long time goal. Apparently, during this initial stage, our aim is to raise many open questions than to actually answer ones. For the following stage, we will realize the 3D power-map using real data in the power grids. One wonders if this new direction will be far-reaching in years to come toward the age of Big Data.

\bibliographystyle{IEEEtran}
\bibliography{helx}

%
%
%
%
%
%
%
%
%
%
%
%
\end{document}